\documentclass[letterpaper]{article} 

\usepackage{aaai2026}  
\nocopyright
\usepackage{amsthm}
\usepackage{algorithm}
\usepackage{algorithmic}
\usepackage{booktabs}  
\usepackage{threeparttable}
\usepackage{amssymb}
\usepackage{amsmath} 
\usepackage{times}  
\usepackage{helvet}  
\usepackage{courier}  
\usepackage[hyphens]{url}  
\usepackage{graphicx} 
\usepackage{natbib}  
\usepackage{caption} 
\usepackage{newfloat}
\usepackage{listings}
\DeclareCaptionStyle{ruled}{labelfont=normalfont,labelsep=colon,strut=off} 
\floatstyle{ruled}
\newfloat{listing}{tb}{lst}{}
\floatname{listing}{Listing}
\title{DOPA: Stealthy and Generalizable Backdoor Attacks from a Single Client under Challenging Federated Constraints}
\author{
Xuezheng Qin\textsuperscript{1},
Ruwei Huang\textsuperscript{1},
Xiaolong Tang\textsuperscript{1},
Feng Li\textsuperscript{2}
}

\affiliations{
\textsuperscript{1}School of Computer, Electronics and Information, Guangxi University, China\\
\textsuperscript{2}China Mobile Communications Group Guangxi Co., Ltd., China\\
2413393056@st.gxu.edu.cn, ruweih@gxu.edu.cn, 2313301044@st.gxu.edu.cn, lifeng3@gx.chinamobile.com
}

\begin{document}

\maketitle

\begin{abstract}

Federated Learning (FL) is increasingly adopted for privacy-preserving collaborative training, but its decentralized nature makes it particularly susceptible to backdoor attacks. Existing attack methods, however, often rely on idealized assumptions and fail to remain effective under real-world constraints, such as limited attacker control, non-IID data distributions, and the presence of diverse defense mechanisms. To address this gap, we propose DOPA (Divergent Optimization Path Attack), a novel framework that simulates heterogeneous local training dynamics and seeks consensus across divergent optimization trajectories to craft universally effective and stealthy backdoor triggers. By leveraging consistency signals across simulated paths to guide optimization, DOPA overcomes the challenge of heterogeneity-induced instability and achieves practical attack viability under stringent federated constraints.
We validate DOPA on a comprehensive suite of 12 defense strategies, two model architectures (ResNet18/VGG16), two datasets (CIFAR-10/TinyImageNet), and both mild and extreme non-IID settings. Despite operating under a single-client, black-box, and sparsely participating threat model, DOPA consistently achieves high attack success, minimal accuracy degradation, low runtime, and long-term persistence. These results demonstrate a more practical attack paradigm, offering new perspectives for designing robust defense strategies in federated learning systems.

\end{abstract}

\begin{links}
\end{links}

\section{Introduction}

Federated Learning (FL)~\citep{FL,Federated—Learning,How-to-backdoor-federated-learning,2016federated}, as a distributed collaborative machine learning paradigm that preserves data privacy, has demonstrated significant potential in critical domains such as healthcare and finance~\citep{mobile,smart-healthcare,smart-healthcare-system,shukla2025federated}. However, its open and distributed nature also introduces new security vulnerabilities. Among these, backdoor attacks\cite{How-to-backdoor-federated-learning} pose a particularly severe threat, as they induce targeted misbehavior only on triggered inputs, leaving benign task performance largely unaffected.
Effectively launching backdoor attacks in a federated setting remains challenging, primarily due to the following three factors:

\begin{itemize}
\item Divergence in client-side optimization dynamics: Due to non-IID local data, clients follow highly inconsistent optimization trajectories~\citep{non-iid1,non-iid2}. A successful attack must overcome gradient conflicts and design a universal backdoor that remains effective through aggregation.

\item Robust defense strategies: Without any knowledge of server-side defenses or their configurations, malicious updates must evade a wide range of detection and suppression mechanisms, ensuring both the stealth and persistence of the attack.

\item Limited attack resources: In realistic scenarios, an attacker typically controls only a small number of clients, sometimes only one, and has limited opportunities to participate in federated training. Under such constraints, each injection must be highly efficient, and its effects must persist throughout the training process.

\end{itemize}

Existing methods still face limitations when confronted with the above challenges simultaneously, and remain insufficiently equipped to handle complex, real-world deployments. To overcome this, we focus on leveraging the structural dynamics of federated optimization to build more generalizable and stealthy backdoor attacks. Our main contributions are summarized as follows:

\begin{itemize}

\item We propose a new attack perspective that treats the divergence in client optimization dynamics, induced by non-IID data, not as a hindrance but as a structural feature to be exploited. Building on this idea, we introduce DOPA (Divergent Optimization Path Attack), a framework that simulates heterogeneous optimization trajectories and forges a consensus update direction among them to generate universal backdoor triggers with strong generalization and long-term persistence, without requiring access to real client data.

\item We validate the feasibility and robustness of the attack under highly constrained settings: operating with only a single client, sparse participation, and both mild and extreme non-IID conditions. DOPA is evaluated against a comprehensive suite of 12 defense strategies, two image classification tasks, and two model architectures. It achieves 70–100\% attack success rates in most scenarios, with minimal optimization overhead and strong persistence under low-resource, stealthy threat conditions.

\end{itemize}

\section{Related Work}
\subsection{Traditional Backdoor Attack Schemes}
Traditional distributed backdoor attacks\cite{DBA} and their variants~\citep{FCBA,shadow_attack} typically rely on multiple malicious clients collaboratively uploading model updates embedded with triggers. These methods typically assume that the attacker controls a certain proportion of clients. Such methods have demonstrated good attack performance in relatively high-resource scenarios, but their deployment cost remains substantial.

Neurotoxin~\cite{neurotoxin} aims to enhance the durability of backdoor attacks in federated learning. Its core idea is to project malicious updates onto relatively low-sensitivity parameter subspaces of the model, thereby reducing the risk of being overwritten by benign training. This approach does not require the attacker to continuously control a large number of clients during training, and remains effective even when only a limited number of clients are compromised.

\subsection{Advanced Attack Strategies}
Recent work has introduced more fine-grained backdoor attack strategies to address increasingly complex adversarial settings in federated learning. For example, the study by \citet{BC_layers} investigates enhancing backdoor activation and main-task fidelity by identifying critical model layers and applying targeted perturbations. This approach is typically built upon partial knowledge of the server-side defense mechanisms (e.g., FLAME), which guides the selection of corresponding attack paths such as layerwise poisoning (LP) or label flipping (LF), in order to improve the stability and stealthiness of the attack.

Another representative work is A3FL\cite{A3FL}, which adapts trigger patterns through optimization and demonstrates strong performance under mildly heterogeneous data distributions, further advancing the development of low-resource backdoor attacks.

\subsection{Defensive Strategies Against Backdoor Attacks}
Over the years, various defense strategies have been proposed to safeguard federated learning (FL) systems against backdoor attacks~\citep{trimmed_mean,rfa,krum,flame,zeno,foolsgold,fltrust,feddf,dp}. Many focus on constraining potentially malicious updates through techniques like gradient clipping, parameter smoothing, or anomaly detection based on client behavior. Typical approaches such as norm-bound filtering, adaptive noise injection, or client-level scoring aim to suppress abnormal updates from influencing the global model.

Another line of work adopts robust aggregation methods, which filter suspicious updates based on metrics like distance, deviation, or frequency. Algorithms such as Krum, Median, and Trimmed Mean discard outliers, while more advanced schemes like FLAME or FoolsGold leverage historical gradients, similarity patterns, or weighting mechanisms to mitigate adversarial impact.

While these defenses have demonstrated commendable performance under standard threat models, their effectiveness often depends on implicit assumptions, such as the presence of multiple malicious clients, noticeable deviations in poisoned updates, or stable client participation patterns. These assumptions are well-suited to threat models involving a large number of compromised clients, where abnormal signals are statistically prominent. However, in practice, the number and behavior of malicious clients are typically unknown. As a result, overly aggressive defense configurations may misclassify benign clients as adversaries, particularly when facing subtle attacks originating from a single client, thereby harming the overall utility of the federated system. Conversely, overly lenient settings may fail to detect stealthy manipulations altogether. This trade-off highlights the challenge of designing defenses that remain effective across a wide range of threat scenarios, particularly in settings characterized by limited prior knowledge and dynamic optimization behaviors.

\section{Preliminaries and Threat Model}
To precisely describe our attack framework, this section first provides a formal definition of federated learning and the associated backdoor threat. Based on this foundation, we construct a threat model that closely reflects real-world complexity and presents significant challenges. This model serves as the basis for the design and evaluation of our proposed method.
\subsection{Federated Learning}
We consider a federated learning system composed of \textbf{N} clients $C_1, ..., C_N$ and a central server $S$, with the goal of collaboratively training a global model $f_\theta$. In the standard \textbf{FedAvg} process, the server distributes the global model $\theta_t$ to a subset of clients at round $t$. Each selected client performs local training and uploads an update $\Delta\theta_t^{(k)}$. The server then aggregates these updates using weighted averaging:

\begin{equation}
\begin{aligned}
\theta_{t+1} = \theta_t + \sum_{k \in S_t} \frac{n_k}{n} \Delta\theta_t^{(k)}
\end{aligned}
\label{eq:fedavg-update}
\end{equation}

where $n_k$ is the local dataset size of client $k$, and $n = \sum_{k \in S_t} n_k$.

A federated backdoor attack aims to implant a hidden malicious task. A successful attack must satisfy two properties:\\
(1) \textbf{Effectiveness}: a poisoned input $x'$, embedded with a trigger $\delta$, is misclassified as the target label $y_t$, i.e., $f_\theta(x') = y_t$;\\
(2) \textbf{Stealthiness}: the attack should not significantly degrade the performance on the main (benign) task.
\subsection{Threat Model}
We construct a threat model that simulates an attacker with minimal capabilities operating under highly constrained training conditions. This threat model serves as the foundation for our experimental design and evaluation, and is composed of three components: the attacker's objective, capability boundaries, and the system environment.
\subsubsection{Attacker's Goal}
The attacker aims to implant a backdoor into the global model under severe training constraints, such that the backdoor exhibits both high stealth and long-term persistence. Specifically, the backdoor should only be activated when the trigger input is present, and must remain functional over time, despite the influence of benign updates and deployed defense mechanisms.

To implement the attack, we follow a standard data poisoning procedure. For each selected local round, the attacker first applies the optimized trigger $\delta$ and mask $M$ to a randomly sampled subset of clean local data, generating poisoned inputs $x' = (1 - M) \odot x + M \odot \delta$. These inputs are relabeled to the target class $y_t$ to form a poisoned dataset. The local model is then trained on this poisoned set for a small number of epochs (e.g., 5), and the resulting update $\Delta\theta_{\text{mal}}$ is uploaded to the server. This process is repeated every time the attacker client is selected, progressively embedding the backdoor into the global model.

\subsubsection{Attacker Capabilities}
\begin{itemize}
    \item \textbf{Control Scope}: The attacker controls only a single client, denoted as \textbf{M = 1}, where \textbf{M} represents the number of malicious clients under the attacker's control. The attacker has full access to its local data, training process, and the content of its uploaded model updates $\Delta\theta_{\text{mal}}$~\cite{How-to-backdoor-federated-learning}.

  \item \textbf{Knowledge Assumptions}: The attacker operates under a strict black-box setting. They can only observe the global model broadcast $f(\theta_t)$ during communication rounds, and have no knowledge of the server-side defense mechanisms, their configurations, or the updates from other clients\citep{Concepts-taxonomy}.
\end{itemize}

Under this setting, the attack strategy must exhibit general evasiveness against unknown defenses, rather than relying on specific prior knowledge—emphasizing adaptability under realistic deployment constraints.

\subsubsection{System Environment}
\begin{itemize}
  \item \textbf{Diverse Heterogeneous Environments}: We assume that client data distributions exhibit varying degrees of non-independent and identically distributed (non-IID) characteristics. We simulate client data heterogeneity using a Dirichlet distribution\cite{Dirichlet} parameterized by $\alpha$. This concentration parameter controls how label distributions are split across clients: a small $\alpha$ (e.g., 0.1) leads to extreme non-IID distributions (where clients mostly have samples from few classes), while a larger $\alpha$ (e.g., 0.9) produces milder non-IID settings with more balanced class distributions.

  \item \textbf{Sparse Participation}: In our setup, only 10\% of clients are randomly selected to participate in training during each round. Moreover, the attack window is constrained to 50 communication rounds, implying that a single malicious client is expected to be selected only 5 times throughout the entire attack phase. The extremely sparse participation greatly restricts both the frequency and total volume of backdoor injections, thereby imposing strict requirements on the efficiency and long-term persistence of each attempt~\citep{Concepts-taxonomy}.

\end{itemize}

\section{Methodology }
\subsection{Motivation and Method Overview}
\subsubsection{Limitations of Existing Attack Paradigms}

As outlined in the introduction, existing federated backdoor attacks struggle when confronted with realistic constraints, such as limited client control, non-IID data, and unknown defense strategies.
We argue that such limitations stem not only from the strengthening of defense mechanisms, but more critically from the fact that current attack paradigms have yet to fully adapt to the dynamic and heterogeneous nature of federated optimization, which ultimately undermines the quality, stability, and generalizability of backdoor injection.

\subsubsection{Design Motivation: From Global View to Optimization Paths}

Most existing backdoor attacks~\citep{A3FL,BC_layers,DBA,F3BA,FCBA,neurotoxin}, whether static or adaptive, treat the federated system as a single evolving entity, typically represented by the global model. They aim to optimize triggers that remain effective across successive global updates. While intuitive, this perspective overlooks a key structural characteristic of federated learning: the global model is merely a temporary aggregation of heterogeneous client updates at each communication round. As a result, triggers tailored to one global model often fail to generalize, facing gradient conflicts~\citep{continual2,continual3} and gradually losing effectiveness as training progresses~\citep{Continual1}.

\subsubsection{Overview of DOPA Framework}

The key to sustaining attack effectiveness in real-world deployments lies not only in enhancing robustness at individual aggregation points, but more importantly in adapting to the dynamic environment shaped by diverse optimization paths. To this end, we propose DOPA (Divergent Optimization Path Attack), a framework that constructs a set of simulated client optimization paths on the attacker side to explicitly model the dynamic heterogeneity inherent in federated systems. By jointly optimizing across these paths, DOPA generates a backdoor template capable of stable activation under various optimization dynamics. This process is entirely based on local simulation, enhancing the backdoor's adaptability to unknown client behaviors and randomized aggregation mechanisms, and thereby strengthening its long-term effectiveness. To enhance backdoor persistence under benign training, we introduce FedFusion, a trigger shaping mechanism that stabilizes initialization and suppresses early anomalies. See Appendix for details.

\subsection{Core Mechanism of the DOPA Framework}

To implement the aforementioned multi-path generalization strategy, our framework first constructs a set of simulated optimization environments on the attacker's side. This approach is designed to find a universally effective trigger by explicitly modeling the dynamic heterogeneity inherent in federated systems. Unlike traditional approaches that simulate non-IID settings merely through data partitioning, our method models heterogeneity from the optimization perspective, capturing more realistic client-specific training paths.

Our methodology is grounded in optimization dynamics theory \citep{SGD}, which models client-side training as a noise-driven process. The divergence in client updates stems from two primary factors: the statistical structure of their local data (reflected in the gradient noise covariance $C_k$) and their specific training configurations (such as the learning rate $\eta_k$). We posit that a truly generalizable trigger must be robust not to a single model, but to this entire spectrum of update dynamics.

\subsubsection{The DOPA Objective: Generalization Across Simulated Paths}

The DOPA framework is explicitly designed to leverage this principle. First, to generate a universally effective trigger, we simulate the heterogeneous federated environment. This is achieved by constructing a set $\mathcal{F}$ of $K$ reference models, each representing a potential client-side optimization path. To ensure divergence among paths, each model is initialized with the global model but then calibrated on a distinct subset of local data and assigned a unique learning rate $\eta_k$, sampled from a range controlled by a heterogeneity factor $\beta$ (see Appendix for details).

Given this simulated environment $\mathcal{F}$, our objective is to optimize a trigger $\delta$ that minimizes the expected attack loss across all these diverse paths. This goal is formalized by the following objective function:
\begin{equation}
\begin{aligned}
\min_\delta \quad & \mathbb{E}_{f_k \sim \mathcal{F}} \Big[
\mathcal{L}_{\text{atk}}\big( \delta;\ f_k,\ D_{\text{sub}} \big)
\Big]
\end{aligned}
\label{eq:attack-objective}
\end{equation}
where $f_k$ is the $k$-th reference model and the path-specific attack loss $\mathcal{L}_{\text{atk}}$ is the standard cross-entropy loss for the backdoor task:
\begin{align}
\mathcal{L}_{\text{atk}}(\delta;\ f_k,\ D_{\text{sub}}) =\ 
& \mathbb{E}_{(x, y) \sim D_{\text{sub}}} \Big[
\mathcal{L}_{\text{CE}}\Big(
f_k\big((1 - M) \odot x \notag \\
& \qquad +\ M \odot \delta\big),\ y_t
\Big)
\Big]
\label{eq:attack-loss}
\end{align}

Here, \( \mathcal{L}_{\text{CE}} \) is the standard cross-entropy loss, \( y_t \) is the target label, \( M \) is the predefined trigger mask, and \( \odot \) denotes element-wise multiplication. The expectation is taken over a small held-out subset \( D_{\text{sub}} \subset D \), which both reflects the gradient noise inherent in federated optimization and reduces computational overhead. This loss encourages the model \( f_k \) to misclassify the trigger-injected input as the target class.

\subsubsection{Solving the Objective via the Gradient Consensus Mechanism}

To solve the objective in Equation~\ref{eq:attack-objective}, simple gradient descent is insufficient, as it does not explicitly leverage the geometric relationship between the different optimization paths. Our goal is to find a shared descent direction that is effective across all paths. To achieve this, we introduce a novel \emph{Gradient Consensus Mechanism} that dynamically modulates the optimization process to efficiently find a robust consensus solution. This mechanism consists of three steps at each iteration:

First, for each of the $K$ reference models, we compute the gradient of the attack loss with respect to the trigger:

\begin{equation}
g_k = \nabla_\delta \mathcal{L}_{\text{atk}}(\delta; f_k, D_{\text{sub}})
\label{eq:grad-def}
\end{equation}

We then aggregate the gradients into a unified update direction, denoted as \( g_{\text{agg}} = \sum_{k=1}^{K} w_k g_k \), where each weight \( w_k \) reflects the relative attack efficacy of the corresponding optimization path. Specifically, paths where the trigger currently induces stronger misclassification are assigned larger weights, promoting gradients that contribute more directly to the backdoor objective. This adaptive weighting scheme emphasizes effective update directions while attenuating those that are misaligned or less indicative of successful backdoor activation.

Second, we quantify the directional alignment among the individual gradients by computing the \emph{Gradient Consensus Factor}, denoted as $C$. A high value of $C$ indicates that gradients are well aligned, suggesting a more generalizable update direction.

\begin{equation}
C = \frac{2}{K(K-1)} \sum_{i=1}^{K-1} \sum_{j=i+1}^{K} \max\left(0, \frac{g_i \cdot g_j}{\|g_i\|_2 \|g_j\|_2}\right)
\label{eq:consensus-factor}
\end{equation}
This factor acts as a measure of confidence in the current update direction. When multiple, diverse optimization paths all agree on how to change the trigger $\delta$, it provides a strong signal that this direction is not an artifact of a single data batch or learning rate, but a truly generalizable improvement.

The $\max(0, \cdot)$ operator ensures only positive alignment contributes, preventing acceleration when paths conflict.

\begin{figure}[t]
\centering
\begin{minipage}[t]{0.3\linewidth}
    \centering
    \includegraphics[width=\linewidth]{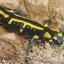}
    \small (a) Clean input
\end{minipage}
\begin{minipage}[t]{0.3\linewidth}
    \centering
    \includegraphics[width=\linewidth]{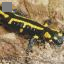}
    \vspace{2pt}
    \small (b) Naive unoptimized trigger
\end{minipage}
\begin{minipage}[t]{0.3\linewidth}
    \centering
    \includegraphics[width=\linewidth]{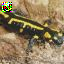}
    \vspace{2pt}
    \small (c) Optimized trigger
\end{minipage}
\caption{Evolution of the Trigger through DOPA Optimization: (a) Clean input, (b) Naive trigger, (c) DOPA-optimized trigger.}
\label{fig:trigger}
\end{figure}

Finally, the magnitude of the update step is modulated by the consensus factor $C$. Strong consensus indicates a highly reliable update direction, justifying an accelerated learning rate. The trigger update is thus given by:
\begin{equation}
\delta \leftarrow \delta - \eta_{\text{eff}} \cdot g_{\text{agg}}, \quad \text{where} \quad \eta_{\text{eff}} = \eta_\delta \cdot (1 + \lambda C)
\label{eq:adaptive-update}
\end{equation}
Here, $\eta_{\text{eff}}$ is the effective learning rate, $\eta_\delta$ is the base rate, and $\lambda$ is a hyperparameter scaling the acceleration sensitivity. This mechanism drives the optimization towards a shared basin of attraction in the optimization landscape, thereby ensuring the trigger's generalizability and robustness.
The entire process is detailed in Algorithm~\ref{alg:dopa}.
We visualize the evolution of the optimized trigger in Figure~\ref{fig:trigger}, illustrating the transformation from a naive pattern to a high-performing backdoor under our framework.


\begin{algorithm}[htb]
\caption{DOPA Trigger Generation}
\label{alg:dopa}
\begin{algorithmic}[1]
\REQUIRE \( \theta_0, \mathcal{D}, y_t, M, K, \eta_0, \beta, \eta_\delta, \lambda, E_{\text{sim}}, E_\delta \)
\STATE Initialize trigger \( \delta \) randomly
\STATE //--- Generate Reference Model Set ---//
\STATE \( \mathcal{F} \leftarrow \emptyset \)
\STATE Generate heterogeneous learning rates \( \{ \eta_k \}_{k=1}^{K} \) parameterized by \( \eta_0, \beta \)
\FOR{ \( k = 1 \) to \( K \) }
    \STATE \( \theta_k \leftarrow \theta_0 \)
    \STATE Sample a data subset \( \mathcal{D}_k \subset \mathcal{D} \)
    \STATE Update \( \theta_k \) for \( E_{\text{sim}} \) epochs on \( \mathcal{D}_k \) using \( \eta_k \)
    \STATE Add \( f_{\theta_k} \) to \( \mathcal{F} \)
\ENDFOR
\STATE //--- Adaptive Optimization via Gradient Consensus ---//
\FOR{ \( e = 1 \) to \( E_\delta \) }
    \STATE Sample a mini-batch \( B \subset \mathcal{D} \)
    \STATE \( \{g_k\}_{k=1}^K \leftarrow \) Compute gradients for \( \delta \) on each \( f_{\theta_k} \in \mathcal{F} \) using \( B \)
    \STATE \( C \leftarrow \) Compute gradient consensus factor from \( \{g_k\} \)
    \STATE \( w_k \leftarrow \) Compute weight for each gradient based on ASR
    \STATE \( g_{\text{agg}} \leftarrow \sum_k w_k \cdot g_k \)
    \STATE \( \eta_{\text{eff}} \leftarrow \eta_\delta \cdot (1 + \lambda C) \)
    \STATE \( \delta \leftarrow \delta - \eta_{\text{eff}} \cdot g_{\text{agg}} \)
\ENDFOR
\RETURN \( \delta \)
\end{algorithmic}
\end{algorithm}

Algorithm \ref{alg:dopa} provides a detailed description of the full process for generating the backdoor trigger $\delta$ under the DOPA framework. The overall procedure is divided into two stages:

\begin{itemize}

 \item In the first stage, a set of reference models $\mathcal{F}$ is constructed to simulate the heterogeneous federated environment. The framework creates $K$ replicas of the global model $\theta_0$ and fine-tunes each for $E_{\text{sim}}$ epochs on a distinct subset of the local data with a unique learning rate, governed by the heterogeneity parameter $\beta$. This process yields a diverse ensemble of models, each representing a potential client optimization path.

 \item The second stage involves optimizing the trigger $\delta$ for $E_\delta$ epochs using the \textbf{Gradient Consensus Mechanism}. In each iteration, gradients of the attack loss with respect to the trigger are computed in parallel across all reference models. These gradients are then used to calculate a consensus factor $C$ and an aggregated gradient $g_{\text{agg}}$, which is weighted by the attack success on each path. The trigger is then updated using an adaptive learning rate $\eta_{\text{eff}}$, which is dynamically accelerated by the consensus factor, ensuring the final trigger is robust across all simulated paths.

\end{itemize}

\section{Experimental Results and Analysis}

{
\setlength{\tabcolsep}{4pt}
\renewcommand{\arraystretch}{1.1}

\begin{table*}[h]
\centering
\begin{threeparttable}

\begin{tabular}{lcccccccccccccc}
\toprule
& FA & NC$_{1.0}$ & NC$_{0.1}$ & FSG & Krum & Med & TM & DP & FedDF & RFA & FLT & MANC & Flame & Zeno \\
\midrule
NTX \textnormal{\scriptsize M=1} & 0.76 & 0.52 & 0.43 & 0.48 & 0.41 & 0.46 & 0.44 & 0.51 & 0.70 & 0.47 & 0.40 & 0.50 & 0.49 & 0.47 \\
NTX \textnormal{\scriptsize M=10} & 63.50 & 3.35 & 0.53 & 64.83 & 0.56 & 0.60 & 1.08 & 1.67 & 69.19 & 0.54 & 0.43 & 0.68 & 0.48 & 0.55 \\
A3FL \textnormal{\scriptsize M=1} & 99.20 & 99.24 & 23.69 & 88.67 & 11.18 & 13.42 & 23.22 & 97.46 & 88.24 & 27.71 & 25.28 & 95.95 & 8.13 & 8.83 \\
A3FL \textnormal{\scriptsize M=5} & \textbf{100} & \textbf{100} & 79.36 & \textbf{100} & 7.85 & 46.03 & 87.00 & 100 & 98.35 & 92.76 & \textbf{96.16} & \textbf{100} & 17.86 & 98.52 \\
Ours \textnormal{\scriptsize M=1} & \textbf{100} & \textbf{100} & \textbf{90.64} & \textbf{100} & \textbf{42.75} & \textbf{99.87} & \textbf{100} & \textbf{100} & \textbf{99.51} & \textbf{100} & 46.02 & \textbf{100} & \textbf{99.12} & \textbf{100} \\
\bottomrule
\end{tabular}
\caption{Attack Success Rate (ASR, \%) of Neurotoxin, A3FL, and Our Method under Mildly non-IID Data ($\alpha{=}0.9$).}
\label{tab:aggregation}
\begin{tablenotes}
\item \textit{Note:} NC$_{x}$ indicates the Norm Clipping defense with clipping threshold $x$. The best result for each column is highlighted in bold.
\end{tablenotes}

\end{threeparttable}
\end{table*}
}

{
\setlength{\tabcolsep}{5pt}
\begin{table*}[h]
\centering
\begin{tabular}{lccccccccccccc}
\toprule
& FA & NC$_{1.0}$ & FSG & Krum & Med & TM & DP & FedDF & RFA & FLT & MANC & Flame & Zeno \\
\midrule
NTX & 1.05 & 0.62 & 0.86 & 1.71 & 0.84 & 0.68 & 0.61 & 0.94 & 0.53 & 0.98 & 0.49 & 0.61 & 0.63 \\
A3FL & 17.02 & 13.26 & 36.62 & \textbf{70.62} & 1.28 & 1.35 & 15.43 & 9.03 & 1.66 & 2.61 & 3.14 & 11.82 & 35.54 \\
Ours & \textbf{99.99} & \textbf{100} & \textbf{99.97} & 46.86 & \textbf{76.29} & \textbf{99.95} & \textbf{100.00} & \textbf{95.34} & \textbf{100.00} & \textbf{37.88} & \textbf{99.99} & \textbf{99.98} & \textbf{100.00} \\
\bottomrule
\end{tabular}
\caption{Attack Success Rate (ASR, \%) Comparison under Extreme non-IID Data ($\alpha{=}0.1$).}
\begin{tablenotes}
\item \textit{Note:} Under a more challenging setting with extremely non-IID client data ($\alpha{=}0.1$), our method maintains relatively stable ASR across most defenses. All results are reported under M=1. The best result for each column is highlighted in bold.
\end{tablenotes}
\label{tab:cifar10_0.1}
\end{table*}
}

\subsection{Experimental Setup}
To comprehensively and rigorously evaluate our attack framework, we design a strict experimental protocol with the following components:

\subsubsection{Scenarios and Models}
Experiments are conducted on the CIFAR-10\cite{cifar10} and TinyImageNet\cite{tiny-imagenet} datasets\cite{dataset—jieshao}, using ResNet18\cite{resnet18} and VGG16\cite{vgg16} as backbone models. We simulate a federated network with 100 clients, where 10 clients are randomly selected per communication round. Tests are performed under two levels of data heterogeneity: mild non-IID ($\alpha = 0.9$) and extreme non-IID ($\alpha = 0.1$).All attacks are launched from a clean pre-trained model, within the first 50 communication rounds. Attack Success Rate (ASR) is measured as the mean over communication rounds 51–60. Extended evaluation on VGG16 and TinyImageNet is provided in the Appendix. Further ablation studies on key design choices, such as the heterogeneity factor $\beta$ and the number of reference models \( K \), are also included in the Appendix.

\subsubsection{Evaluation Metrics}
We use Attack Success Rate (ASR, \%) to measure effectiveness, and the change in Main Task Accuracy (MTA), denoted as $\Delta\mathrm{MTA}$, to evaluate stealthiness. We further assess attack overhead by recording trigger optimization time, and quantify long-term persistence by tracking ASR over extended communication rounds~\citep{dataset—jieshao,badnets}.

\subsubsection{Baselines and Defenses}
We compare our method (Ours, $M = 1$) with two representative baselines: Neurotoxin (NTX, early-stage) and A3FL (SOTA). The baselines are evaluated at $M = 1, 5, 10$. To assess robustness, we construct a comprehensive evaluation suite comprising 13 aggregation strategies, including 11 mainstream defenses from prior literature, the standard FedAvg baseline, and a novel adaptive variant we designed. These strategies cover robust aggregation~\citep{trimmed_mean,rfa}, update filtering~\citep{krum,  foolsgold, fltrust, zeno}, norm clipping and privacy-preserving methods~\citep{dp}, as well as hybrid or learning-based defenses~\citep{feddf, flame}. For brevity, we use abbreviations for all strategies in the main text; full names and descriptions are provided in the Appendix.

\subsubsection{Fairness Assurance}
All experiments are conducted using a unified defense codebase. For each baseline, we search for the optimal hyperparameter configuration to reproduce the original reported results, ensuring maximum fairness and reproducibility.

\subsection{Robustness under Severe Data Heterogeneity}

\begin{figure}[t]
\centering
\includegraphics[width=0.70\linewidth]{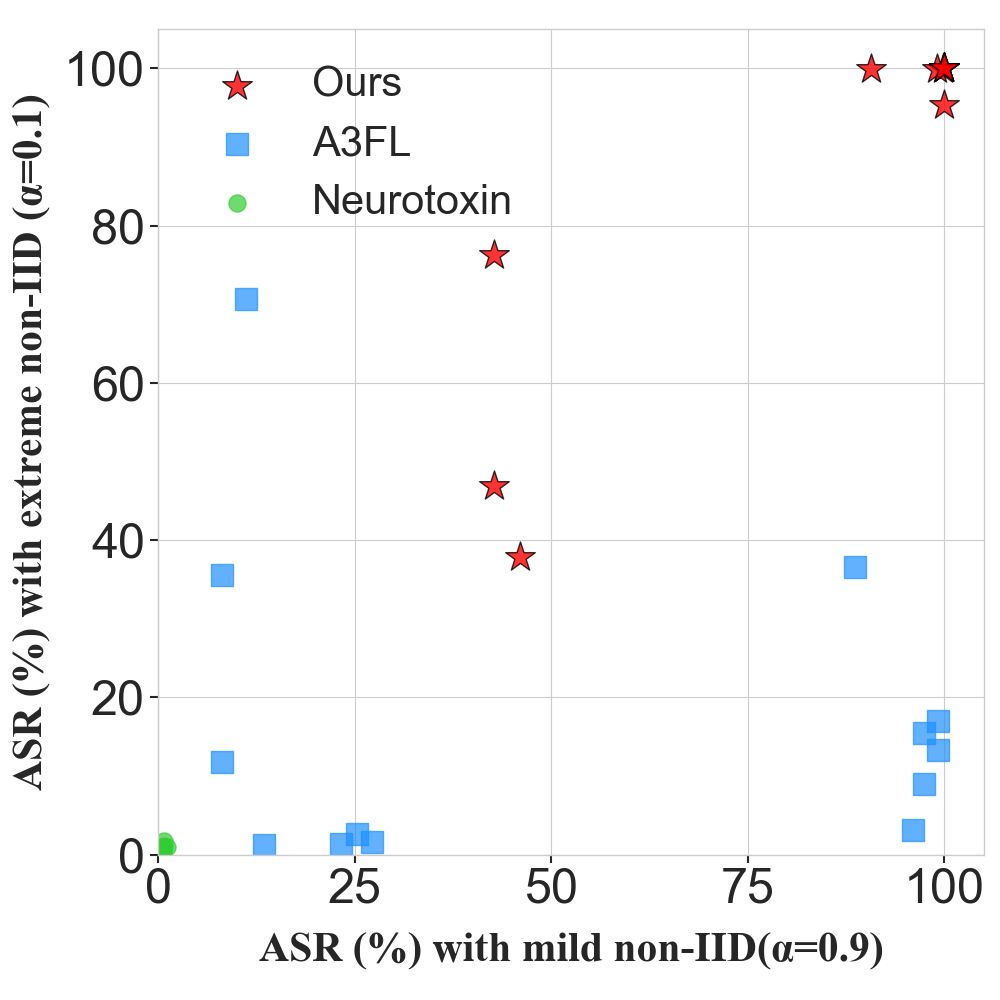}
\caption{Attack robustness to data heterogeneity on CIFAR-10 ($M{=}1$ ). ASR (\%) under mild non-IID ($\alpha{=}0.9$) vs. extreme non-IID ($\alpha{=}0.1$) settings.}
\label{fig:asr_heterogeneity}
\end{figure}

Figure~\ref{fig:asr_heterogeneity} compares the attack success rates (ASR) of various defenses under different data heterogeneity levels. Each marker represents a defense method under backdoor attack. Our method maintains consistently high ASR, indicating strong robustness across heterogeneity levels.

\begin{table}[t]
\centering
\begin{threeparttable}
\begin{tabular}{lccc}
\toprule
& Ours & A3FL & Neurotoxin \\
\midrule
CIFAR-10 & 61s & 179s & 141s \\
Tiny-ImageNet & 153s & 782s & 420s \\
\bottomrule
\end{tabular}
\caption{Comparison of one-round trigger optimization overhead (seconds). Results are measured under mildly non-IID data ($\alpha = 0.9$).}
\label{tab:trigger_optimization_time}

\end{threeparttable}
\end{table}

\subsection{Attack Effectiveness under Diverse Defense Strategies}
We evaluate the effectiveness of our method under various defense strategies in both mildly non-IID ($\alpha = 0.9$) and extremely non-IID ($\alpha = 0.1$) settings, and compare it against two representative baselines: \textbf{A3FL} and \textbf{Neurotoxin}.

\begin{figure*}[t]
\centering
\includegraphics[width=0.88\linewidth
]{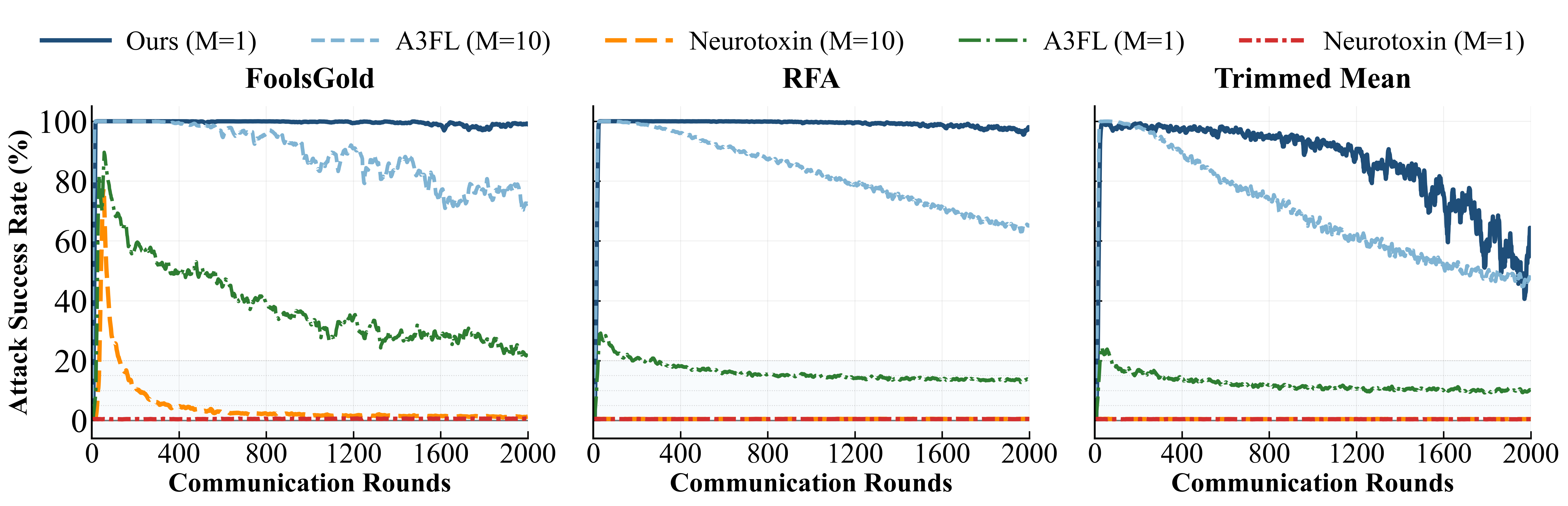}

\caption{
Attack persistence over 2000 rounds. Despite using only one client, our method maintains high ASR, while baselines degrade rapidly despite using more malicious clients. The attack window is the first 50 rounds. 
}

\label{fig:2}
\end{figure*}

{
\setlength{\tabcolsep}{5pt}

\begin{table*}[t]
\centering
\begin{threeparttable}
\setlength{\tabcolsep}{4pt}
\begin{tabular}{lccccccccccccc}
\toprule
& FA & NC & FSG & Krum & Med & TM & DP & FedDF & RFA & FLT & MANC & Flame & Zeno \\
\midrule
Cifar10 & +0.10 & +0.03 & +0.07 & -0.04 & +0.06 & +0.03 & -3.48 & +0.05 & +0.03 & +0.20 & +0.03 & +0.03 & +1.66 \\
Tiny-ImageNet & -0.18 & -0.10 & -0.21 & -0.96 & +0.06 & -0.25 & +1.03 & -0.25 & -0.16 & -0.75 & +0.02 & -0.42 & -0.14 \\
\bottomrule
\end{tabular}
\begin{tablenotes}
\caption{Net Impact of DOPA on Average MTA Over 50 Training Rounds ($\Delta$MTA \%)}
\label{tab:mta_impact}
\item \textit{Note:} This table reports the percentage change in average Main Task Accuracy (MTA) over the 50-round attack window. Positive values indicate a slight improvement, while negative values indicate a slight drop.
\end{tablenotes}
\end{threeparttable}
\end{table*}
}

\subsection{Attack Effectiveness under Mildly non-IID Conditions}

Table~\ref{tab:aggregation} presents the core results under the mildly non-IID distribution. Our method, using only a single malicious client ($M = 1$), achieves ASR comparable to or better than \textbf{A3FL ($M = 5$)} under most defense mechanisms, demonstrating relatively high resource efficiency.

The effectiveness of \textbf{Neurotoxin} is limited under low-resource settings, requiring longer injection windows and more malicious clients. We further evaluate its performance under extended injection windows in the Appendix and observe results similar to those reported in the original paper.

As shown in Table~\ref{tab:cifar10_0.1}, even under extreme data heterogeneity ($\alpha = 0.1$), our attack achieves over 70\% ASR against 10 out of 12 defense methods, indicating its robustness and adaptability in highly skewed data distributions.

\subsection{Computational Efficiency}

Our method achieves effective backdoor injection while exhibiting improved computational efficiency. As shown in Table~\ref{tab:trigger_optimization_time}, the trigger optimization process is approximately 2.9× and 2.3× faster than A3FL and Neurotoxin, respectively, on CIFAR-10. On the more complex TinyImageNet dataset, the speedup increases to roughly 5.1× and 2.8×.

This efficiency arises from the gradient consensus mechanism introduced in DOPA. While prior approaches often optimize within a single global context without modeling diverse training trajectories, DOPA simulates multiple optimization paths and aggregates their gradient signals in parallel to identify a unified update direction with strong alignment. When consistent directional trends emerge, the framework reinforces the update step, allowing the trigger to converge rapidly toward a generalizable solution. As a result, the optimization becomes more streamlined and requires fewer iterations to reach high attack efficacy, significantly reducing total computational cost.

\subsection{Attack Persistence Over Training Rounds}
We further evaluate the persistence of the attack, which serves as a key indicator of its practical threat level in long-term federated training. As shown in Figure~\ref{fig:2}, our method (Ours, M=1) demonstrates stable and sustained attack performance under various defense mechanisms:

\begin{itemize}
\item Under several mainstream defenses (e.g., Foolsgold, RFA), the ASR remains close to 100\% throughout 2000 communication rounds.

\item 
Even under stronger defenses such as Trimmed Mean, the attack remains effective for up to 1200 rounds.

\item 
In contrast, previously proposed methods, even when using more malicious clients, do not consistently maintain attack persistence across extended rounds.
\end{itemize}

These results suggest that our framework enables highly persistent backdoor injection, with its effect remaining stable over extended training periods.

\subsection{Stealthiness of the Attack: Impact on Main Task Performance}

Beyond attack effectiveness, stealthiness is crucial to evaluating backdoor threats. We quantify its long-term impact by the change in average MTA during the injection period, defined as:
\begin{equation}
\Delta \text{MTA} = \text{Avg-MTA}_{\text{w/Attack}} - \text{Avg-MTA}_{\text{w/o Attack}}
\end{equation}

We conduct a controlled experiment to measure this impact. Starting from the same pre-trained model, we compute its average MTA over 50 training rounds in a benign federated learning environment as the baseline ($\text{Avg-MTA}_{\text{w/o Attack}}$). Under identical settings, we then apply the DOPA attack and measure the average MTA during the 50-round injection period ($\text{Avg-MTA}_{\text{w/Attack}}$). All evaluations are performed under mild data heterogeneity ($\alpha=0.9$).

Experimental results are summarized in Table~\ref{tab:mta_impact}. The data show that DOPA causes only minor disruptions to main task performance. Across most defense scenarios on both CIFAR-10 and Tiny-ImageNet, MTA fluctuations remain within approximately $\pm 0.5\%$ throughout training. This consistently low impact highlights the strong stealthiness of DOPA, allowing malicious behavior to persist with minimal degradation to benign performance.

\subsection{Conclusion and Outlook}
We present DOPA, a backdoor attack framework that remains highly effective under single-client, black-box, and non-IID federated settings. By simulating heterogeneous optimization dynamics, DOPA achieves strong attack success, long-term persistence, and minimal accuracy degradation across diverse defenses.

We conclude by highlighting a promising yet challenging direction for future research. Like most existing backdoor attack methods, our approach relies on an attacker-defined trigger. However, in real-world deployments, such explicit triggers are unlikely to appear naturally unless the attacker can actively manipulate the input stream. Future work may explore more naturalistic attack vectors and extend DOPA-style path simulation mechanisms to trigger-free settings, particularly in natural language processing and multimodal tasks. Such directions could move beyond current paradigms and uncover more stealthy and realistic threat models.

\clearpage

\appendix 
\section{Derivation of the Variance Aggregation Formula}
\subsection{Gradient Noise Modeling Assumption}
We begin from the perspective of a single client. In stochastic gradient descent (SGD), client k computes a stochastic gradient using a mini-batch of data as follows:

\begin{equation}
g_t^{(k)}(\theta_t) = \nabla L_k(\theta_t) + \varepsilon_t^{(k)}
\label{eq:stochastic_gradient}
\end{equation}

where the gradient noise $\varepsilon_t^{(k)} \sim \mathcal{N}(0, C_k)$ represents the mathematical manifestation of inherent optimization heterogeneity. This noise arises from the randomness of mini-batch sampling, and is a key reason why different clients, even with the same learning rate, may follow distinct optimization paths. The covariance matrix $C_k$ reflects the characteristics of the local data on each client.

\paragraph{Assumption 1 (Single-Step Deterministic Gradient Assumption)}
To simplify subsequent derivations, we introduce the following assumption:

\begin{quote}
\textit{Within a single local update step, the true gradient $\nabla L_k(\theta_t)$ is treated as a deterministic constant. Therefore, all stochasticity in client updates is solely attributed to the mini-batch sampling noise $\varepsilon_t^{(k)}$.}
\end{quote}

Substituting Equation~\ref{eq:stochastic_gradient} into the SGD update rule:

\begin{equation}
\theta_{t+1} = \theta_t - \eta_k\, g_t^{(k)}(\theta_t)
\label{eq:sgd_update}
\end{equation}

we obtain:

\begin{equation}
\theta_{t+1} - \theta_t = -\eta_k\, \nabla L_k(\theta_t) - \eta_k\, \varepsilon_t^{(k)}
\label{eq:sgd_update_expanded}
\end{equation}

\subsection{Quantifying the Variance of Local Updates}
Next, we quantify this stochasticity, which arises from two components, by analyzing the variance of the local model update vector:

\begin{equation}
\Delta \theta^{(k)} = -\eta_k\, g_t^{(k)}(\theta_t)
\label{eq:local_update_vector}
\end{equation}

Using the properties of variance and applying Assumption 1, which treats $\nabla L_k(\theta_t)$ as deterministic, we arrive at the following critical relationship:

\begin{equation}
\operatorname{Var}(\Delta \theta^{(k)}) = \eta_k^2\, \operatorname{Var}(g_t^{(k)}(\theta_t)) = \eta_k^2\, C_k
\label{eq:local_update_variance}
\end{equation}

The term $\eta_k^2\, C_k$ characterizes the variance structure of the local update and captures the intensity of perturbations during the optimization process.

\subsection{Global Statistical Characteristics of Aggregated Updates}
Finally, we shift our perspective to the server side. Assume the server aggregates updates from $K$ clients, each statistically independent. Then, the variance of the global model update $\overline{\Delta \theta}$ is given by:

\begin{equation}
\operatorname{Var}(\overline{\Delta \theta}) = \operatorname{Var} \left( \frac{1}{K} \sum_{k=1}^{K} \Delta \theta^{(k)} \right) = \frac{1}{K^2} \sum_{k=1}^{K} \operatorname{Var}(\Delta \theta^{(k)})
\label{eq:global_update_variance_step1}
\end{equation}

Substituting the conclusion from Equation~\ref{eq:local_update_variance}, we obtain the final formula:

\begin{equation}
\operatorname{Var}(\overline{\Delta \theta}) = \frac{1}{K^2} \sum_{k=1}^{K} \eta_k^2\, C_k
\label{eq:global_update_variance_final}
\end{equation}

\subsection{Implications and Applications}
This derivation quantitatively illustrates the relationship between local optimization behaviors and the global statistical properties of model aggregation. It provides theoretical support for the heterogeneous simulation mechanism employed by the DOPA framework under resource-constrained attack scenarios. Compared to traditional methods that rely on data partitioning, this approach offers greater flexibility and controllability, making it well-suited for backdoor attacks in black-box settings.

\section{Experimental Setup}

This section provides the full details of the experimental setup referenced in the main text.

\subsection{Defense Methods and Strategies}

In our experiments, we evaluate a total of 13 aggregation methods. This includes the standard federated averaging algorithm (FedAvg), eleven mainstream backdoor defense strategies published in the literature, and an additional baseline defense we constructed and implemented, which we refer to as \textit{Median-Anchored Norm Clipping} (MANC), designed to rigorously test the robustness of our attack against adaptive clipping mechanisms.

\subsubsection{Standard Federated Averaging Algorithm}

\textbf{FedAvg (Baseline Method)} is the standard federated averaging (FA) algorithm in which the server updates the global model by computing a weighted average of client-submitted updates. It contains no built-in defense mechanisms and serves as the primary baseline for evaluating the effectiveness of all defense strategies.

\subsubsection{Defensive Aggregation Strategies}

To assess robustness, we construct a comprehensive evaluation suite comprising 13 aggregation strategies, including 11 mainstream defenses from prior literature, the standard FedAvg baseline, and a novel adaptive variant we designed. These strategies cover robust aggregation, update filtering, norm clipping and privacy-preserving methods, as well as hybrid or learning-based defenses.

\paragraph{Robust Aggregation}

\begin{itemize}
    \item \textbf{Median (Med) / Trimmed Mean (TM):} Two classic \textit{coordinate-wise} defenses. For every model dimension they independently compute either the median or the mean after trimming extreme values, giving them strong resistance against large, abnormal parameter values injected by attackers. We use a trim ratio $\beta = 0.2$ for Trimmed Mean (i.e., discarding the largest and smallest 20\% values along each coordinate).

    \item \textbf{Robust Federated Aggregation (RFA):} Another \textit{distance-based} defense that iteratively computes the geometric median of all client updates—replacing the traditional mean—to reduce the influence of spatial outliers. We run 10 iterations of Weiszfeld's algorithm per round with a convergence threshold $\varepsilon = 10^{-5}$.
\end{itemize}

\paragraph{Update Filtering}

\begin{itemize}

    \item \textbf{Krum:} A \textit{distance-based} defense that measures the distance between each client update and all others, and selects the one whose update is most consistent with its $n - f - 2$ nearest neighbors (i.e., has the smallest sum of distances to them). We set the maximum tolerated number of Byzantine clients to $f = 2$ per round, and follow the original constraint that requires at least $n \ge 2f + 3$ clients to perform Krum safely.

    \item \textbf{Foolsgold (FSG):} A \textit{similarity-based} defense that calculates the cosine similarity between update vectors to identify and down-weight ``colluding'' updates that are highly aligned in direction.

    \item \textbf{FLTrust (FLT):} A \textit{similarity-based correction} method that computes the directional similarity between each client's update and a trusted server-side reference update, then performs weighted aggregation accordingly to penalize anomalous directions.

    \item \textbf{Zeno:} A unique \textit{suspicion-score-based filtering mechanism}. It scores each client update based on its estimated contribution to loss reduction on a server-side validation set, penalized by the update's gradient norm. Before each aggregation round, Zeno removes the $b$ updates with the lowest scores, assuming they are most likely to be adversarial. We adopt the original hyperparameters $n_r = 4,\; \rho = 0.0005$, and set the filtering threshold to $b = 1$, which aligns with the empirical upper bound for 10\% malicious clients. This setting ensures robustness while avoiding excessive removal of benign updates.

\end{itemize}

\paragraph{Norm Clipping and Privacy-Preserving Methods}

\begin{itemize}
    \item \textbf{Norm Clipping (NC) and Differential Privacy (DP):} Two strategies based on \textit{norm clipping and perturbation}. The former imposes a fixed upper bound on the $L_2$ norm of each update; the latter adds statistical noise on top of clipping, which enhances resistance to precisely crafted backdoor injections while also protecting privacy. In our experiments, we test NC with clipping thresholds of $1.0$ and $0.1$, while the DP setting uses a fixed clipping bound of $1.0$ and adds Gaussian noise with standard deviation $\sigma = 0.0015$.
\end{itemize}

\paragraph{Hybrid or Learning-Based Defenses}

\begin{itemize}
    \item \textbf{MANC:} A hybrid and adaptive defense that dynamically suppresses updates that deviate too far from a robust anchor point constructed from the coordinate-wise median.

    \item \textbf{FedDF:} A unique \textit{knowledge-distillation-based} defense that bypasses direct aggregation entirely. Instead, the server distills knowledge from all client models into a new global model using a shared public dataset.

    \item \textbf{Flame:} A hybrid and adaptive defense that leverages clustering and adaptive clipping. For FLAME, we set the maximum tolerated number of adversaries to $f=2$ (i.e., HDBSCAN uses cluster size $n - f$), and inject Gaussian noise with $\lambda = 0.001$ per round.
\end{itemize}

\subsection{Implementation Details \& Fairness}

{
\setlength{\tabcolsep}{5pt}
\begin{table*}[h]
\centering
\begin{threeparttable}
\begin{tabular}{lccccccccccccc}
\toprule
\textbf{Method} & FA & NC & FSG & Krum & Med & TM & DP & FedDF & RFA & FLT & MANC & Flame & Zeno \\
\midrule
NTX & 0.26 & 0.20 & 0.21 & 0.42 & 0.18 & 0.24 & 0.14 & 0.36 & 0.15 & 0.32 & 0.21 & 0.22 & 0.28 \\
A3FL & 58.21 & 41.29 & 57.83 & \textbf{ 1.90} & 4.06 & 8.67 & 41.30 & 48.40 & 18.15 & 87.17 & 14.43 & 1.56 & 33.36 \\
Ours & \textbf{77.98} & \textbf{90.45} & \textbf{99.45} & 0.88 & \textbf{20.27} & \textbf{54.06} & \textbf{91.31} & \textbf{96.60} & \textbf{83.59} & \textbf{98.97} & \textbf{50.69} & \textbf{1.69} & \textbf{98.71} \\
\bottomrule
\end{tabular}
\caption{Generalization to Complex Datasets: ASR (\%) on TinyImageNet with ResNet18 ($\alpha=0.9$)}
\label{tab:tinyimagenet}
\begin{tablenotes}
\item \textit{Note:} This table shows the ASR on the more complex TinyImageNet dataset. All results are reported under M=1. The best result for each column is highlighted in bold.
\end{tablenotes}
\end{threeparttable}
\end{table*}
}

{
\setlength{\tabcolsep}{5pt}
\renewcommand{\arraystretch}{1.1}
\begin{table*}[h]
\centering
\begin{threeparttable}
\begin{tabular}{lccccccccccccc}
\toprule
\textbf{Method} & FA & NC & FSG & Krum & Med & TM & DP & FedDF & RFA & FLT & MANC & Flame & Zeno \\
\midrule
NTX & 1.86 & 1.00 & 0.93 & 0.89 & 0.92 & 0.94 & 1.00 & 0.84 & 0.93 & 0.93 & 0.96 & 0.96 & 0.91 \\
A3FL & 99.13 & 98.95 & 36.79 & 37.69 & 36.51 & 40.30 & 96.49 & 60.43 & 39.96 & 49.59 & 96.80 & 99.47 & 96.34 \\
Ours & \textbf{100} & \textbf{99.97} & \textbf{99.87} & \textbf{98.53} & \textbf{99.59} & \textbf{99.76} & \textbf{100} & \textbf{99.22} & \textbf{99.75} & \textbf{99.81} & \textbf{99.99} & \textbf{100} & \textbf{99.61} \\
\bottomrule
\end{tabular}
\caption{Generalization to Different Architectures: ASR (\%) on CIFAR-10 with VGG16 ($\alpha=0.9$)}
\label{tab:vgg16}
\begin{tablenotes}
\item \textit{Note:} This table reports the ASR when the model architecture is switched to VGG16. Our method maintains consistently high attack success rates (close to 100\%) across all 12 defenses, indicating robust adaptability to architectural changes.All results are reported under M=1. The best result for each column is highlighted in bold.
\end{tablenotes}
\end{threeparttable}
\end{table*}
}

To ensure fairness and reproducibility of our experimental results, we adhere to the following rigorous principles:

\paragraph{Attack Parameters.} To ensure fair comparison, we adopt the same backdoor trigger configuration as all baseline methods. Specifically, the trigger is consistently placed in the top-left corner of the input image, and its size is adjusted according to the dataset resolution: 5×5 pixels for CIFAR-10 (32×32), and 10×10 pixels for TinyImageNet (64×64). All attack experiments, including baselines and ours, are conducted under a fixed random seed of 20 to ensure reproducibility.

\paragraph{Framework Parameters.} Our DOPA engine simulates 3 reference models in parallel by default. The heterogeneity control parameter $\beta$ is fixed at 0.2 across all experiments. In Appendix \emph{DOPA Framework: Mechanism Analysis}, we demonstrate that the effectiveness of our method is not sensitive to fine-tuning this parameter.

\paragraph{Hardware and Statistics.} All experiments are conducted on a single NVIDIA 4070Ti SUPER GPU using PyTorch.

Through this rigorous and comprehensive experimental design, this section aims to objectively present the real-world threat level of our proposed attack framework in federated learning scenarios.

\section{Generalization and Detailed Analysis}

\subsection{Cross-Dataset and Cross-Architecture Generalization}

To validate the generalizability of our framework, we extend the evaluation to a more complex dataset (Tiny-ImageNet) and a different model architecture.

As shown in the corresponding table, after migrating to TinyImageNet, our method maintains an attack success rate (ASR) above 80\% under more than half (7 out of 12) of the evaluated defense mechanisms, demonstrating stability and adaptability in a more challenging optimization environment.

To further investigate the impact of model architecture on attack performance, we replace the backbone network from ResNet18 to VGG16. Experimental results show that under this architecture, the effectiveness of our method is further enhanced: our attack achieves an average attack success rate (ASR) of 99\% across all 12 defense mechanisms,  marking a further improvement over the ResNet18-based setting. This shows high robustness across model architectures.

We hypothesize that the strong attack performance observed under the VGG16 architecture is closely associated with its \textit{over-parameterized} nature. 
Compared to modern architectures such as ResNet, VGG16 contains a significant amount of parameter redundancy when applied to datasets like CIFAR-10, 
which is generally associated with a \textit{smoother loss landscape}, facilitating optimization. 
We believe that this architectural trait benefits our attack framework in two key ways:

\begin{enumerate}
    \item \textbf{Easier Backdoor Implantation.} 
    The large model capacity provides more redundant parameters, allowing the backdoor trigger to be embedded without significantly disrupting the main task. 
    This improves the stealthiness of the attack.
    
    \item \textbf{Lower Optimization Difficulty.} 
    In a smoother and more expansive parameter space, our DOPA framework converges more efficiently and is able to identify a robust and generalized attack pattern 
    that remains effective across all simulated heterogeneous environments.
\end{enumerate}


\subsection{In-Depth Analysis under Extreme Heterogeneity}


\begin{figure*}[t]
\centering
\begin{minipage}[t]{0.4\linewidth}
    \centering
    \includegraphics[width=\linewidth]{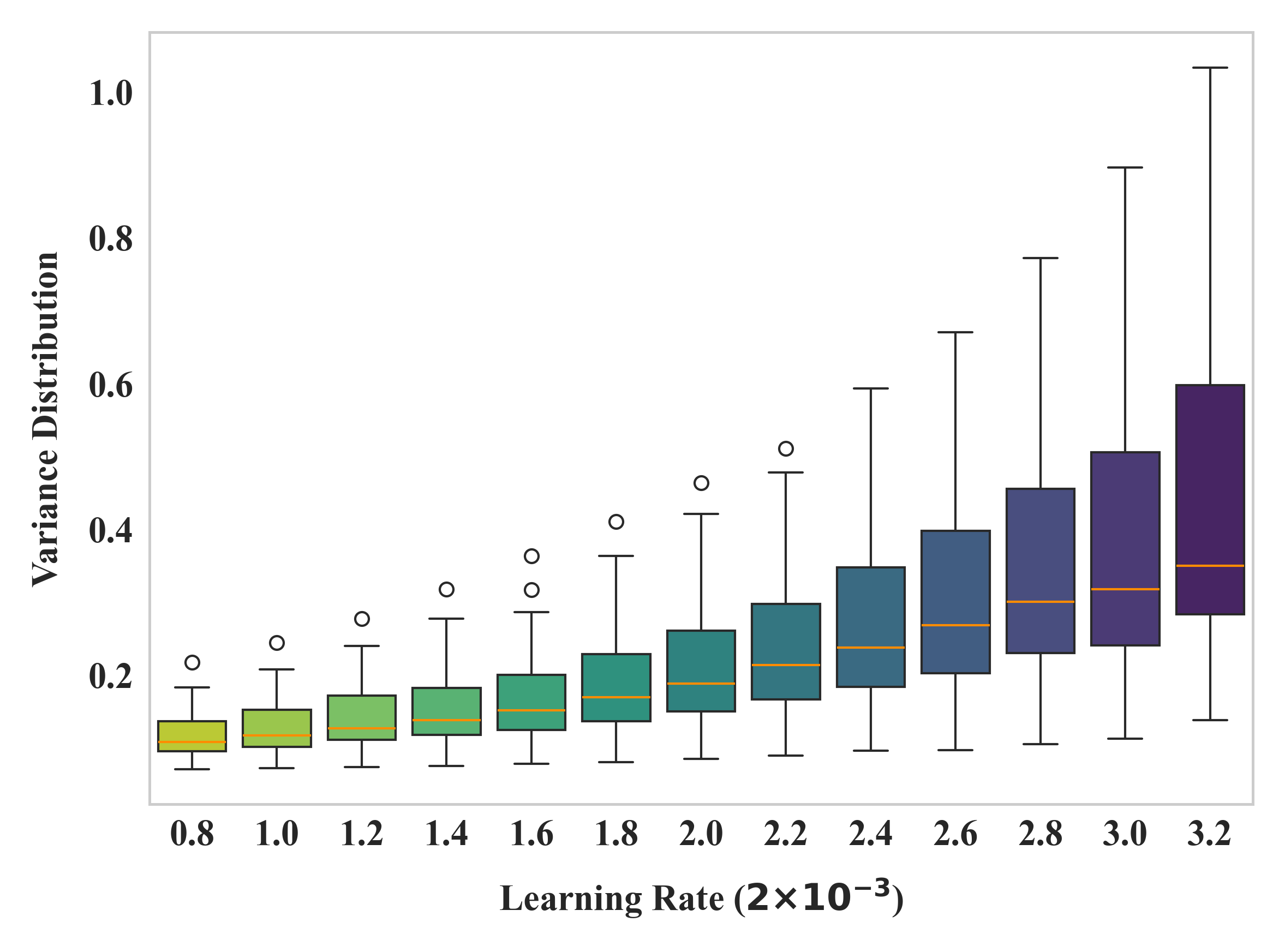}
    \small (a) Variance vs. Learning Rate Scaling Factor ($\gamma$)
\end{minipage}
\begin{minipage}[t]{0.4\linewidth}
    \centering
    \includegraphics[width=\linewidth]{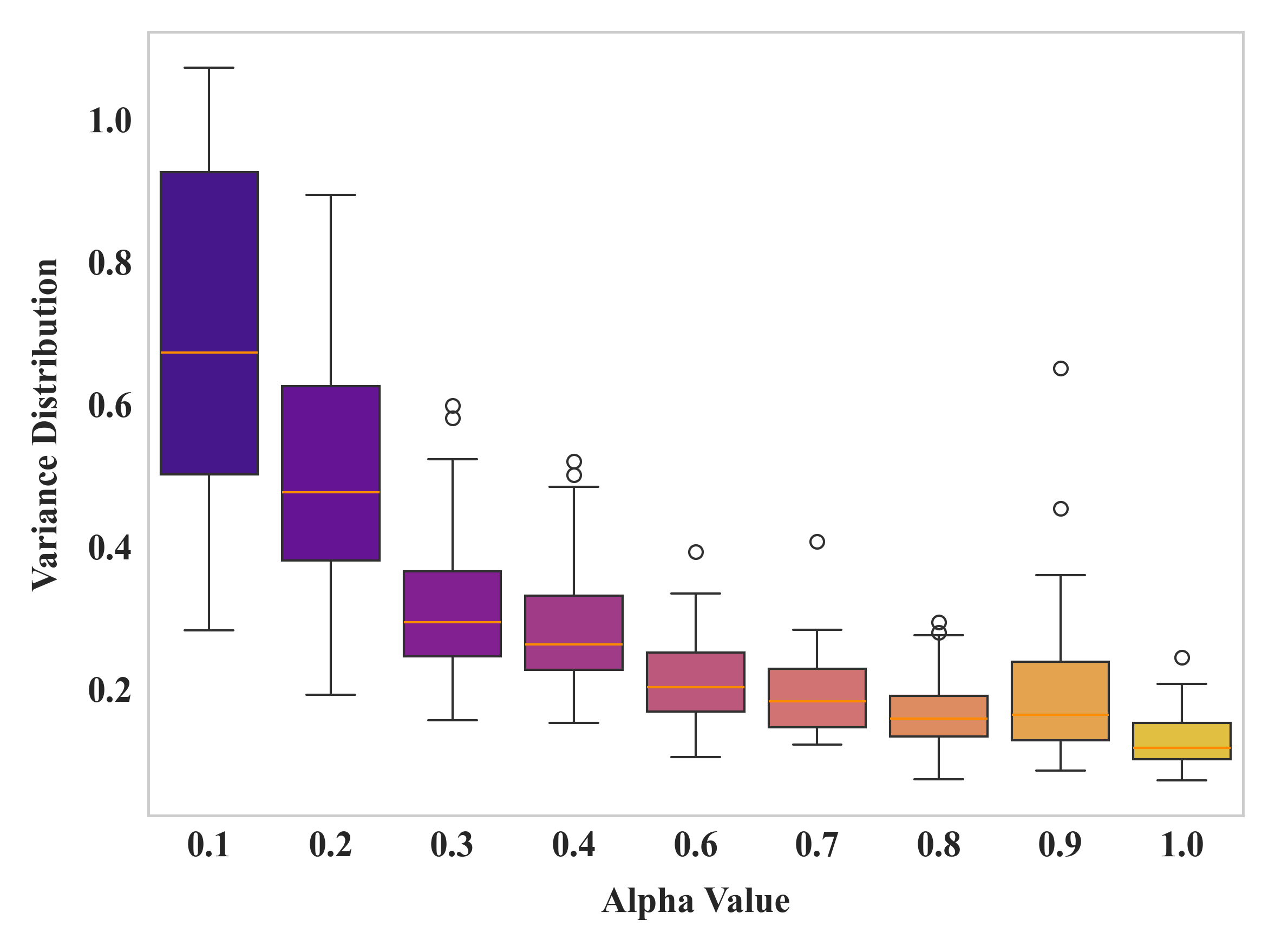}
    \small (b) Variance vs. Data Heterogeneity ($\alpha$)
\end{minipage}
\caption{Statistical equivalence between simulated and real-world heterogeneity. The variance of model updates increases systematically with (a) the learning rate scaling factor $\gamma$ in our simulation and (b) the degree of data heterogeneity (i.e., a smaller Dirichlet parameter $\alpha$). }
\label{fig:heterogeneity_simulation}
\end{figure*}

Table~2 in the main text demonstrates the robustness of our method under extreme data heterogeneity ($\alpha=0.1$). This section provides a deeper analysis of the intricate attack-defense dynamics, with a particular focus on three representative defenses that exhibit markedly different behaviors: the failure of Krum, and the effectiveness of Median and FLTrust.

\paragraph{Krum.} A noteworthy observation is that A3FL's ASR under Krum rises sharply from 11.18\% in mildly heterogeneous settings ($\alpha=0.9$) to 70.62\% under extreme heterogeneity ($\alpha=0.1$). This can be attributed to the core assumption underlying Krum—that benign client updates should form relatively tight clusters in parameter space. However, in highly non-IID data distributions, benign updates themselves become significantly more dispersed, weakening this assumption. Consequently, malicious updates are more likely to be statistically "camouflaged" within the spread of benign ones.

\paragraph{Median.} This defense operates independently on each parameter coordinate. Its effectiveness stems from filtering based on value distributions rather than relying on the geometric location of entire vectors. Under extreme heterogeneity, even though the full update vectors from benign clients may be highly scattered, the updates on certain critical coordinates targeted by attackers typically remain small. As a result, Median effectively eliminates large abnormal values injected at these coordinates, thereby disrupting the attack at a fine-grained level.

\paragraph{FLTrust.} The core of this method lies in evaluating update \textit{directions} via cosine similarity, rather than their absolute positions. While extreme heterogeneity causes benign update directions to vary, they still statistically align toward the main task's optimization objective. In contrast, the gradient direction of the backdoor task is nearly orthogonal to that of the main task, making malicious updates systematic directional outliers. This gives FLTrust strong detection capability based on directional deviation, unaffected by the overall spread of benign updates.

\vspace{0.5em}
Under extreme non-IID settings, Median and FLTrust exhibit a certain degree of resistance to our attack, which indirectly reflects the multidimensional performance differences among defense mechanisms and helps to comprehensively characterize the boundaries of attack effectiveness. Overall, our framework still achieves a high ASR under most defense strategies, demonstrating good adaptability.

\section{DOPA Framework: Mechanism Analysis}

\begin{figure}[t]
\centering
\includegraphics[width=0.95\linewidth]{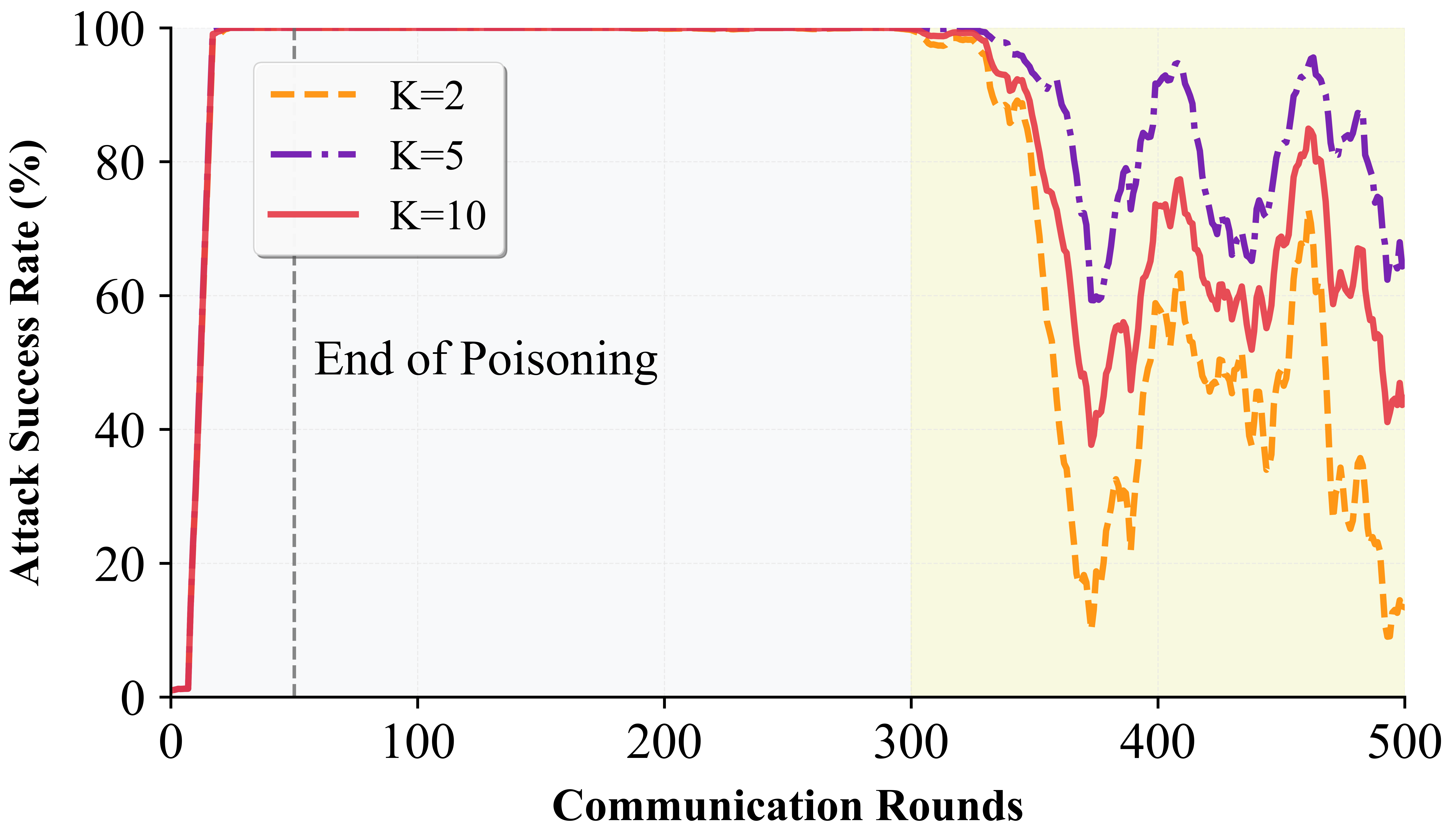}
\caption{Ablation study on the number of reference models $K$. The figure shows the attack success rate across communication rounds under different settings: $K=2$, $K=5$, and $K=10$, with DP defense applied after the poisoning phase.}
\label{fig:ablation_K}

\end{figure}

\begin{figure*}[t]
\centering
\begin{minipage}[t]{0.4\linewidth}
    \centering
    \includegraphics[width=\linewidth]{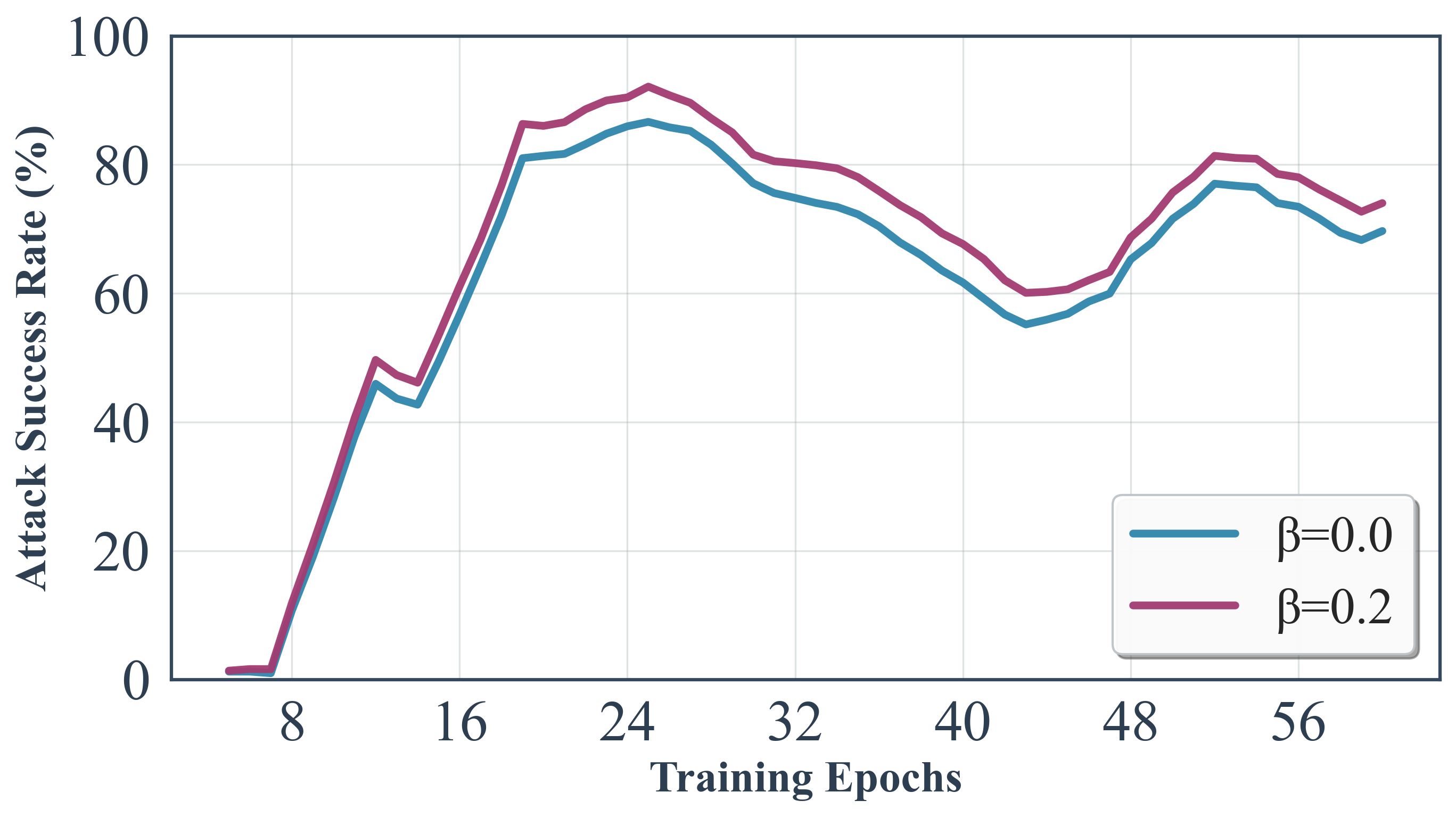}
    \small (a) ASR vs. Training Epochs under Different $\beta$ Settings, smoothed using a window of size 5
\end{minipage}
\begin{minipage}[t]{0.4\linewidth}
    \centering
    \includegraphics[width=\linewidth]{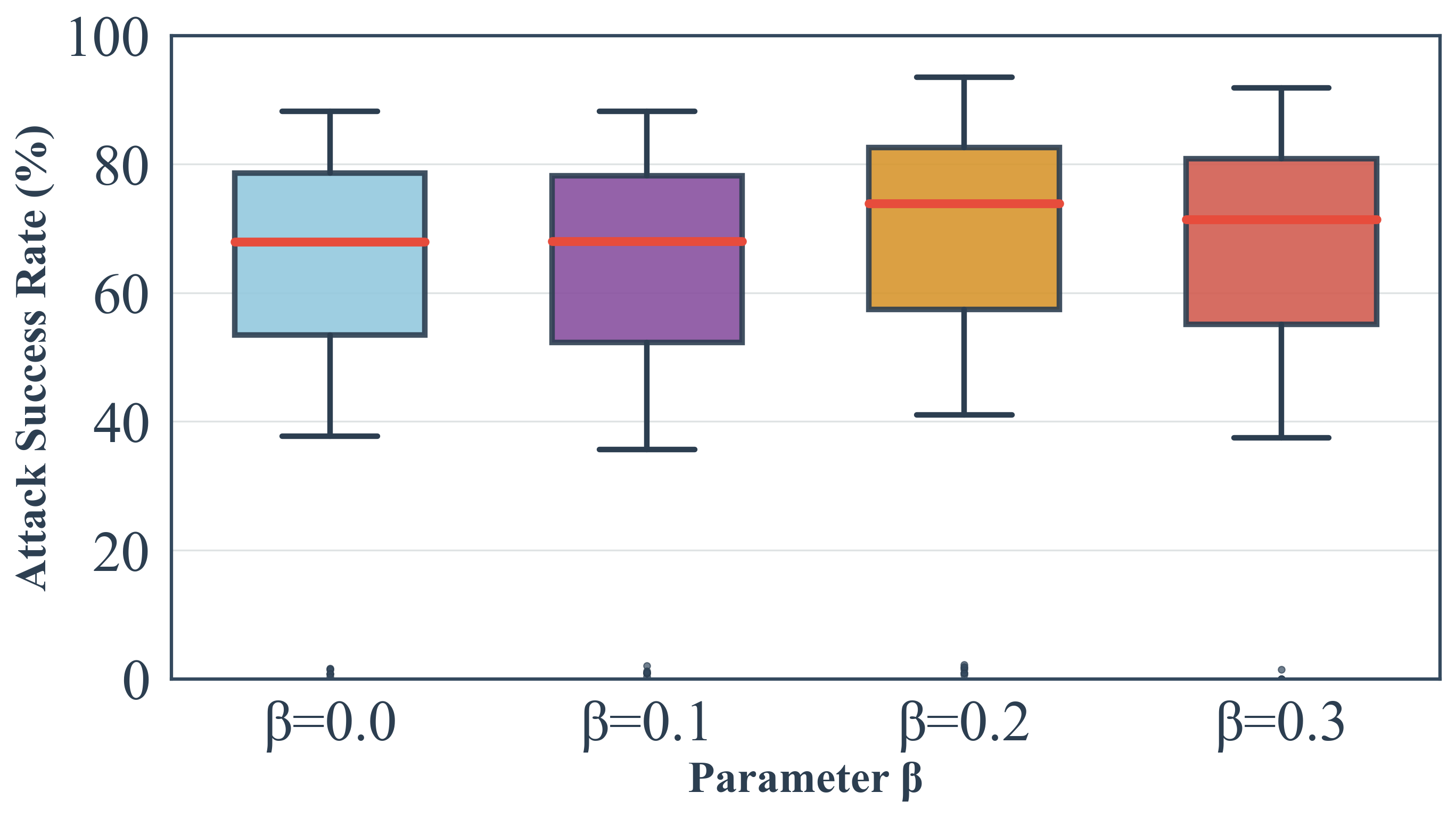}
    \small (b) ASR Distribution under Varying Learning Rate Heterogeneity $\beta$
\end{minipage}
\caption{Sensitivity analysis of hyperparameter $\beta$ in the DOPA framework.}
\label{fig:beta}
\end{figure*}

\subsection{Ablation Study on the Number of Reference Models ($K$)}

To rigorously evaluate the role of path diversity in the DOPA framework, we conducted an ablation study examining how the number of simulated reference models, denoted by $K$, affects the long-term resilience of the backdoor. This experiment was deliberately conducted under a highly adversarial setting using Differential Privacy (DP) as the defense mechanism. Due to its continuous noise injection, DP creates a hostile environment for trigger persistence, making it an ideal candidate for stress-testing the robustness of backdoor attacks.

To further accentuate performance differences across configurations, we also intentionally reduced the number of trigger optimization epochs, generating a sub-optimal trigger. This setup increases the pressure on the attack's inherent robustness, making the effects of varying $K$ more observable.

Figure~\ref{fig:ablation_K} presents the Attack Success Rate (ASR) over time under three configurations: $K=2$, $K=5$, and $K=10$. The results reveal a non-monotonic relationship between $K$ and attack persistence, highlighting a key trade-off in the design of our framework:

\begin{itemize}
    \item \textbf{Insufficient Diversity ($K=2$):} With too few reference models, the simulated optimization trajectories lack adequate heterogeneity. The trigger overfits to a narrow set of training dynamics and fails to generalize, leading to early degradation under DP defense.

    \item \textbf{Optimal Trade-off ($K=5$):} A moderate value achieves the best balance. Sufficient diversity enables robust optimization, while gradient directions remain aligned enough for effective consensus. This setting yields the most stable and persistent attack behavior.

    \item \textbf{Excessive Divergence ($K=10$):} When the number of reference models is further increased to $K=10$, we \textit{hypothesize} that the divergence among simulated optimization paths becomes excessively large, leading to significant gradient conflicts. In this case, the optimizer is forced to reconcile highly inconsistent update directions, which may result in a diluted consensus that fails to strongly align with any specific attack trajectory. As a consequence, the generated trigger becomes less effective, and the attack's overall persistence deteriorates. This suggests that beyond a certain point, increasing $K$ may introduce excessive heterogeneity that hinders rather than helps the backdoor optimization process.
\end{itemize}

In conclusion, the number of reference models $K$ is a crucial hyperparameter that governs the trade-off between diversity and coherence. Our findings suggest that $K=5$ offers a practical and robust default choice for achieving long-term backdoor persistence in federated settings.

\subsection{Validation of the Heterogeneity Simulation Mechanism}

To examine whether the simulation mechanism in DOPA can introduce sufficient diversity among reference models, we conduct a set of controlled experiments. Specifically, we analyze the variance of client model updates induced by varying learning rates, and compare it with the variance resulting from different degrees of data heterogeneity, controlled by the Dirichlet parameter $\alpha$.

Figure~\ref{fig:heterogeneity_simulation}(a) shows that increasing the learning rate scaling factor $\gamma$ leads to a gradual rise in both the median and dispersion of update variance, indicating that larger learning rates introduce greater behavioral differences across clients. Similarly, as shown in Figure~\ref{fig:heterogeneity_simulation}(b), smaller $\alpha$ values (i.e., more heterogeneous data) also result in higher variance. These trends align with the theoretical relationship derived in Appendix \emph{Derivation of the Variance Aggregation Formula}, and suggest that our simulation strategy provides a reasonable approximation of client update diversity in federated settings.

\subsection{Analysis of Key Mechanisms}

We further investigate the stability of DOPA's attack performance by conducting a sensitivity analysis on the hyperparameter $\beta$, which controls the range of learning rate perturbations. To avoid performance saturation from masking marginal effects, we adopt the Median aggregation strategy, known for its robustness under highly heterogeneous settings. As shown in Figure~\ref{fig:beta}, both the temporal ASR trajectories and the statistical distribution across different $\beta$ values remain highly consistent. These results suggest that the DOPA framework is robust and largely insensitive to the choice of $\beta$.

\section{Transferability Analysis of the Optimized Trigger}
\label{sec:transferability}


\begin{table*}[t] 
\centering
\setlength{\tabcolsep}{6pt}
\renewcommand{\arraystretch}{1.2}
\begin{tabular}{ll}
\toprule
\textbf{Component / Metric} & \textbf{Specification / Value} \\
\midrule
Victim Model & Clean ResNet18 (trained via FedAvg) \\
\quad \textit{Benign Task Accuracy} & \textit{92.21\%} \\
\addlinespace 
Attack Trigger & DOPA-optimized on CIFAR-10 \\
\quad \textit{ASR during its optimization} & \textit{45.83\%} \\
\addlinespace
\textbf{Transfer Attack ASR (on Victim)} & \textbf{50.38\%} \\
\bottomrule
\end{tabular}
\caption{Verification of the DOPA trigger's transferability. The test evaluates a trigger, optimized in a separate DOPA process, against a fully benign ResNet18 model that was never exposed to malicious updates.}
\label{tab:transferability}
\end{table*}


\subsection{Unexpected Attack Effectiveness under Full Defense Isolation}
In experiments employing strict aggregation defenses such as \textit{Krum}, we observed an unexpected yet noteworthy phenomenon: in certain runs, the malicious client was never selected for aggregation during the entire attack window (i.e., all malicious updates were successfully filtered by the defense mechanism). Despite this complete isolation, the final global model nonetheless exhibited a non-negligible susceptibility to the trigger generated by our method, with the attack success rate (ASR) reaching 40–50\% under certain conditions. This observation challenges the conventional assumption that successful backdoor attacks necessarily rely on direct parameter injection, and suggests that, even in the absence of explicit model poisoning, a trigger may still exert a meaningful influence.

\subsection{Independent Validation in an Isolated Setting}
To rule out potential confounding factors in the original training pipeline, we designed a fully decoupled evaluation procedure. Specifically, we exported the optimized trigger and mask tensors produced by DOPA and evaluated them in a clean, isolated environment. In this setting, we loaded a benign model trained from scratch using standard FedAvg, without any exposure to poisoned clients or malicious updates. The saved trigger was then applied to this model, and its ASR was measured on the target task.

\subsection{Experimental Results}
As shown in Table~\ref{tab:transferability}, the trigger achieved an \textbf{ASR of 50.38\%} on the fully benign model. Given that this model had never encountered any malicious updates during training, the result indicates that the trigger's effectiveness does not rely on a specific model instance. Instead, it appears capable of activating its intended behavior across similarly structured models, demonstrating a degree of cross-model generalization.

\subsection{Interpretation and Hypothesis}
These findings suggest that the triggers optimized by DOPA exhibit a certain level of transferability across models under specific settings. Unlike traditional backdoor attacks that rely on parameter corruption, these triggers are more likely to exploit structural vulnerabilities inherent in the learning task, shaped jointly by the model architecture and data distribution. We hypothesize that the core mechanism of DOPA, namely joint optimization across multiple simulated, heterogeneous client trajectories, enables it to identify perturbations that are robust across varied training dynamics. As a result, the final trigger behaves more like a \textit{universal adversarial pattern} rather than a tailor-made implant for a specific model instance.

Interestingly, this transferability appears to be task-dependent. In additional experiments, we replaced CIFAR-10 with the more complex Tiny-ImageNet dataset, keeping the model architecture (ResNet18) and the DOPA framework unchanged. Under this new setting, the resulting trigger consistently exhibited low ASR against Krum, failing to reproduce the strong cross-model effect observed on CIFAR-10.

This contrast suggests that the effectiveness of the DOPA-optimized trigger is not universal, but rather depends on the compatibility between the model architecture (e.g., ResNet18) and the data distribution (e.g., CIFAR-10). When task complexity increases (as in Tiny-ImageNet), the corresponding structural vulnerability may become less prominent or more difficult to exploit using standard optimization. This observation provides valuable insight into the boundaries of such attacks and highlights the importance of considering model--task interactions when evaluating backdoor strength and persistence.

\begin{quote}
A more systematic exploration of model--task combinations and vulnerability attribution will be a key focus of our future work.
\end{quote}

\subsection{Potential Implications for Defense Research}
These observations may offer new perspectives for designing federated learning defenses. While most existing strategies focus on identifying and filtering anomalous updates, such approaches may fall short in scenarios where the attack does not rely on explicit parameter corruption.

We hypothesize that such attacks are linked to generalizable vulnerabilities in the input space---manifesting as triggers that can activate undesired behavior even without ever being aggregated. As such, future defense strategies may benefit from incorporating geometric analyses of decision boundaries, enhancing robustness to universal perturbations, or leveraging theoretically grounded certified defenses.

Although further theoretical analysis and empirical validation are still needed, our findings suggest that threats beyond the traditional \textit{inject-and-detect} paradigm may exist---threats that are more subtle, transferable, and equally worthy of attention.

\section{FedFusion Mechanism}

To improve the survivability of the backdoor after the attack window closes and under prolonged rounds of benign training, we propose an auxiliary module \textbf{FedFusion}. Without compromising attack efficacy, it introduces two complementary statistical regularizers that explicitly guide the trigger $\delta$ to evolve towards a \emph{low-amplitude, non-outlier} pattern.

\begin{figure}[h]
\centering
\includegraphics[width=0.8\linewidth]{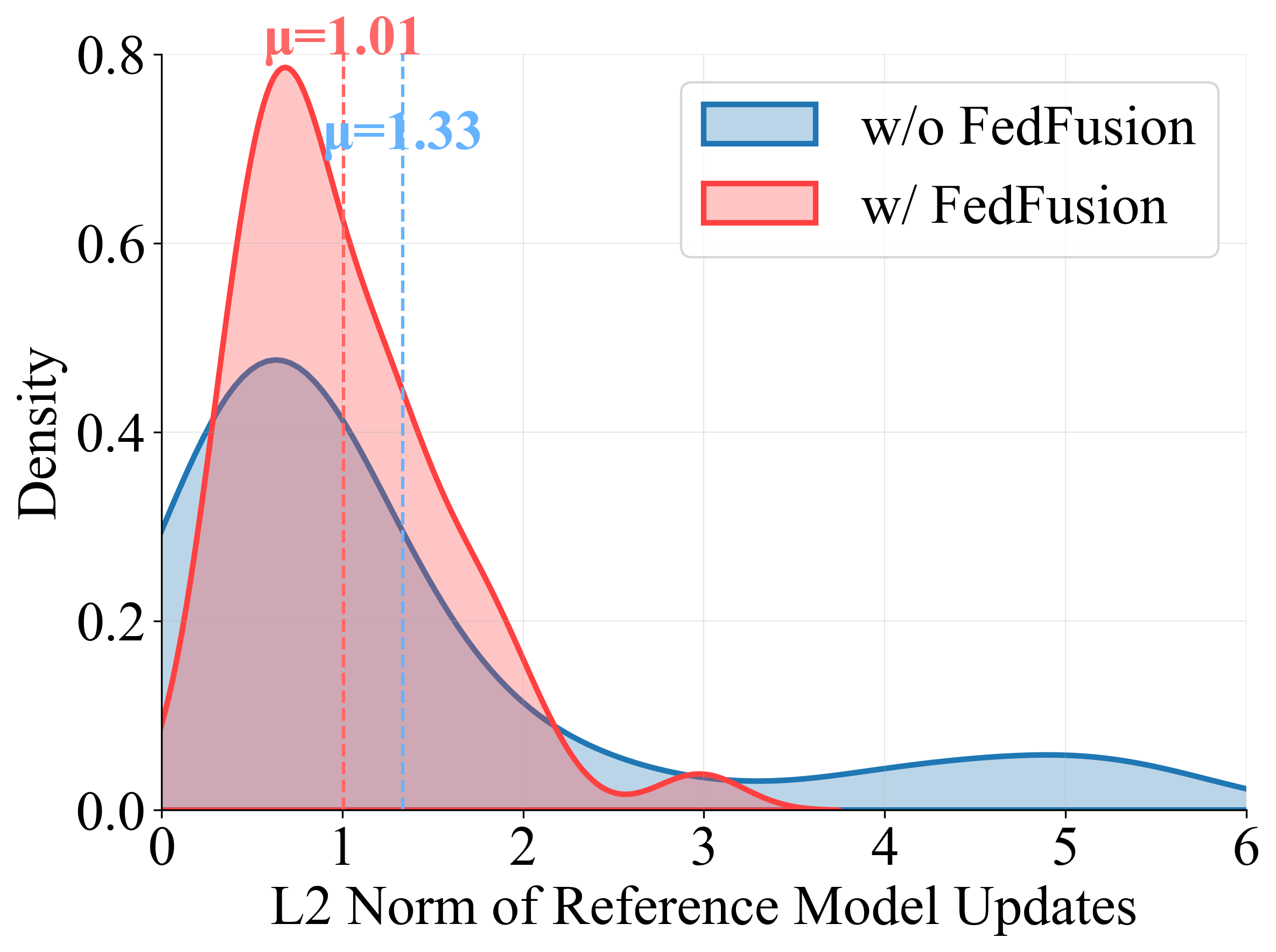}
\caption{Effect of FedFusion on the distribution of reference model update norms. 
The red curve (with FedFusion) exhibits a more concentrated and lower-magnitude update pattern, compared to the more dispersed and spiky distribution without regularization (blue curve).}
\label{fig:fedfusion_distribution}
\end{figure}

\begin{figure*}[t]
\centering
\includegraphics[width=0.90\linewidth
]{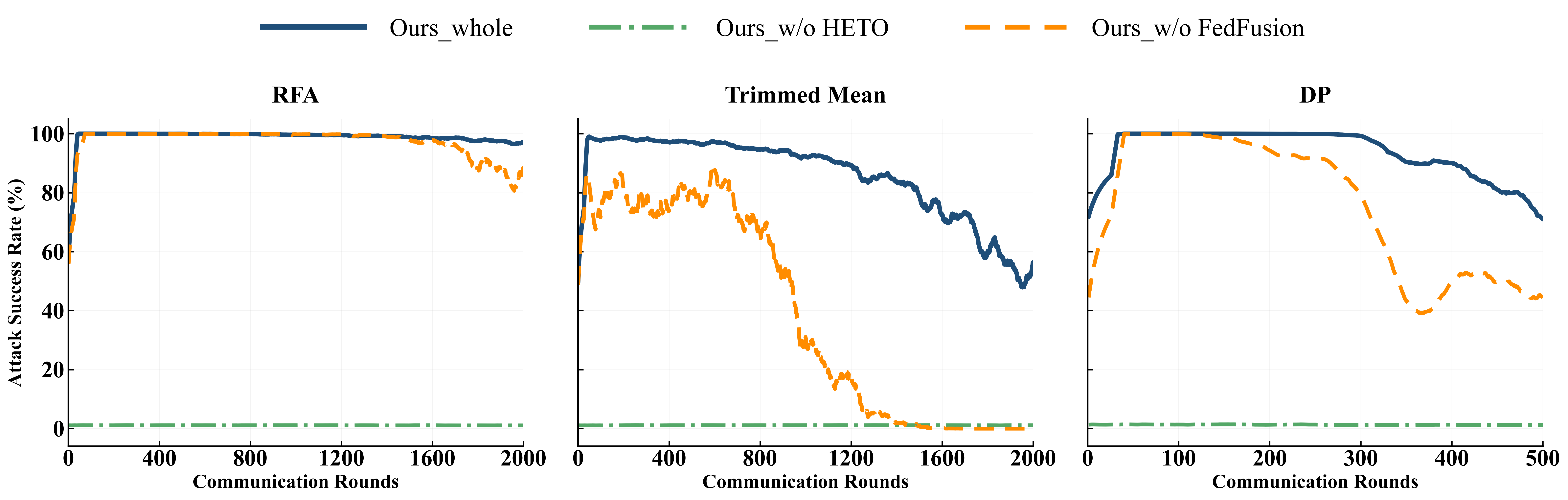}
\caption{
Ablation on attack persistence under RFA, Trimmed Mean, and DP defenses. Each subplot shows the ASR (\%) of our full method (blue solid), w/o FedFusion (orange dashed), and w/o DOPA (green dash-dotted), over 2000 rounds (500 for DP). $M{=}1$ attacker is used. The first 50 rounds are the attack window.
}

\label{fig:ablation}
\end{figure*}

We observe in the internal optimization trajectory of the DOPA engine that, when starting from an unregulated trigger (e.g., a solid color patch), the constructed reference paths in DOPA tend to induce a more dispersed norm distribution, occasionally exhibiting high-magnitude anomalous perturbations (see Figure~\ref{fig:fedfusion_distribution}). Although these reference models are not directly uploaded, their training trajectories influence the final convergence of the trigger $\delta$, leading it to adopt a similarly  \emph{spiky} parameter update style upon real-world injection. Such concentrated, high-amplitude perturbations on a small subset of parameters are more likely to be averaged out during subsequent rounds of benign aggregation, thereby undermining the long-term stealth of the backdoor.

To this end, \textbf{FedFusion} applies the following joint regularization to the trigger $\delta$ before the DOPA optimization stage:

\begin{align}
\mathcal{L}_{\text{fusion}} =\ 
& (\mu(\delta) - \mu_{\text{target}})^2 
+ (\sigma(\delta) - \sigma_{\text{target}})^2 \notag \\
& +\ \text{ReLU}\left( \|\delta\|_2 - C_{\text{budget}} \right)^2
\label{eq:fusion-loss}
\end{align}

where the first two terms (collectively denoted as $\mathcal{L}_{\text{diff}}$) constrain the pixel-wise mean $\mu(\delta)$ and standard deviation $\sigma(\delta)$ of the trigger, encouraging them to statistically resemble the characteristics of a uniform distribution $\mathcal{U}(0,1)$—specifically, $\mu_{\text{target}} = 0.5$ and $\sigma_{\text{target}} = 0.288$. This helps avoid brightness bias and structural monotonicity, promoting the generation of more complex and unstructured trigger patterns. The third term, $\mathcal{L}_{\text{ceil}}$, imposes a relatively strict upper bound $C_{\text{budget}} = 1.0$ on the L2 norm of $\delta$ to suppress potential high-magnitude divergence along the optimization trajectory. This constraint is not meant to mimic any defense threshold, but rather serves as a proactive robustness regularization from the attacker's side, shaping a safer perturbation initialization in the early optimization phase.

With the above regularization applied, the distribution of model updates induced by the trigger exhibits a marked shift in statistical characteristics. As shown by the red curve in Figure~\ref{fig:fedfusion_distribution}, the update norms become significantly more concentrated, and high-magnitude outliers are notably suppressed, resulting in a smoother and more stable perturbation regime. This trend remains consistent across various tasks, indicating that the mechanism possesses strong transferability and attacker-side controllability, thereby effectively enhancing the long-term survivability of the backdoor under complex federated aggregation dynamics.

The values $\mu_{\text{target}} = 0.5$, $\sigma_{\text{target}} = 0.288$, and $C_{\text{budget}} = 1.0$ are fixed hyperparameters and are not specifically tuned for any particular scenario or defense strategy. For a more detailed analysis of FedFusion's standalone contribution within the overall attack framework, please refer to the ablation study in Appendix \emph{Ablation Study}.

\section{Ablation Study}

To precisely dissect the individual contributions of each component within our framework, we conduct a comprehensive ablation study. Taking the full version of our framework (\textit{Ours}) as the baseline, we evaluate the performance impact of removing each core module in turn. Experiments are performed under three representative defense mechanisms: Trimmed Mean, RFA, and DP.

\subsection{DOPA: Core Driver of Attack Behavior}

As shown in the figure~\ref{fig:ablation}, when the DOPA module is removed, the attack success rate (ASR) remains consistently close to zero across all three defense settings, indicating complete failure of the attack. This result highlights that DOPA serves as the foundational mechanism for generating effective backdoor signals in our framework, and its function cannot be substituted by any other component.

\subsection{FedFusion: Auxiliary Module for Stealth Enhancement}

FedFusion is designed to optimize the statistical distribution of the uploaded model updates, thereby improving the survivability of the backdoor. Under the Trimmed Mean and DP defenses, removing FedFusion leads to a notable degradation in attack performance, as reflected in:

\begin{itemize}
    \item \textbf{Trimmed Mean}: The ASR drops rapidly after around 1000 rounds and eventually approaches zero;
    \item \textbf{DP}: The backdoor remains effective in the early phase but experiences a clear ``late-stage collapse'' after approximately 300 rounds.
\end{itemize}

In contrast, under the \textbf{RFA} defense, the impact of FedFusion is relatively minor. This suggests that RFA's distance-based robust aggregation mechanism is not highly sensitive to the statistical properties of parameter distributions (e.g., L2 norm), and thus the stealth enhancement offered by FedFusion is less effective in this setting.

\section{Impact of Attack Window on Neurotoxin}

In our main experiments, all baseline methods were evaluated under a 50-round attack window to ensure consistency and comparability. However, we observed that Neurotoxin's attack success rate was significantly lower than what was reported in the original paper. To investigate this discrepancy, we conducted additional experiments in this section.

\begin{figure}[h]
\centering
\includegraphics[width=0.99\linewidth]{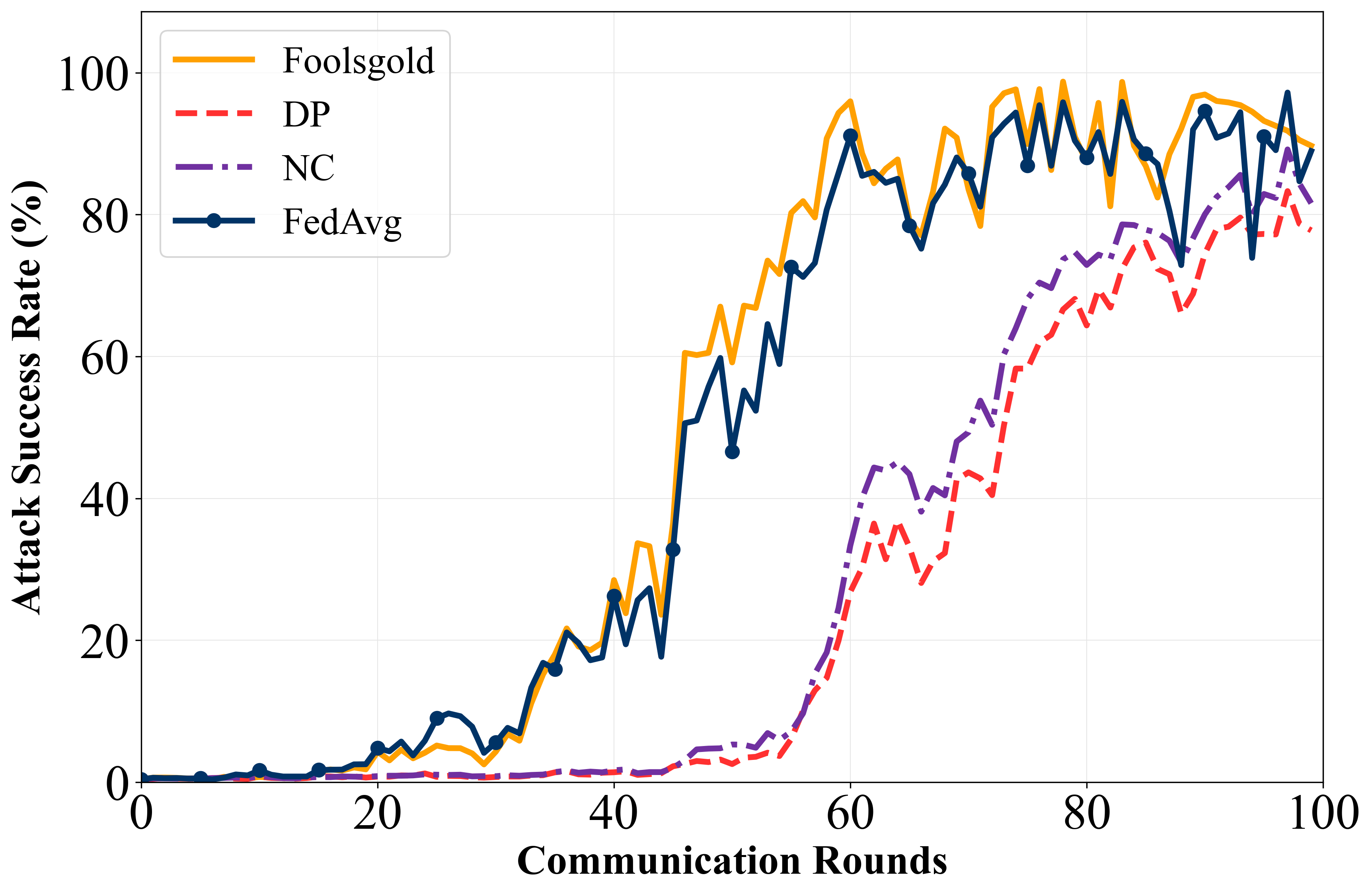}
\caption{Attack success rate (ASR, \%) of Neurotoxin under various defense mechanisms with a 100-round attack window.}
\label{fig:asr_Neurotoxin}
\end{figure}

Neurotoxin relies on continuously injecting low-detectability updates, and its attack effectiveness is largely influenced by the injection frequency and the total number of rounds. In the original paper, the attacker client was selected in 40 consecutive rounds. In contrast, under our setting, the attacker was selected fewer than ten times throughout the entire process, resulting in a lower injection frequency and total volume, which may have limited its effectiveness.

To verify this impact, we extended the attack window to 100 rounds while keeping other parameters constant, increased the number of attackers to M=10, and re-evaluated performance under several defense mechanisms (e.g., FedAvg, Differential Privacy, Clipping, FoolsGold). The results are shown in Figure~\ref{fig:asr_Neurotoxin}.

Experimental results show that extending the attack window significantly improves Neurotoxin's attack success rate (ASR), with noticeable gains observed under multiple defense strategies. In particular, the results under aggregation mechanisms such as clipping and differential privacy (DP) align with the original paper and follow-up works like A3FL, further demonstrating that Neurotoxin's effectiveness is closely tied to the length of the injection window.

In summary, the supplementary experiments in this section indicate that our reproduction of Neurotoxin in the main experiments was correct. The performance gap mainly stems from differences in the attack window setting, rather than implementation errors. This further highlights that, in federated learning, the length of the attack window can have a significant impact on the performance of certain attack methods and should be clearly specified during evaluation.


\begin{thebibliography}{38}
\providecommand{\natexlab}[1]{#1}

\bibitem[{Bagdasaryan et~al.(2020)Bagdasaryan, Veit, Hua, Estrin, and Shmatikov}]{How-to-backdoor-federated-learning}
Bagdasaryan, E.; Veit, A.; Hua, Y.; Estrin, D.; and Shmatikov, V. 2020.
\newblock How to backdoor federated learning.
\newblock In \emph{International conference on artificial intelligence and statistics}, 2938--2948. PMLR.

\bibitem[{Blanchard et~al.(2017)Blanchard, El~Mhamdi, Guerraoui, and Stainer}]{krum}
Blanchard, P.; El~Mhamdi, E.~M.; Guerraoui, R.; and Stainer, J. 2017.
\newblock Machine learning with adversaries: Byzantine tolerant gradient descent.
\newblock \emph{Advances in neural information processing systems}, 30.

\bibitem[{Cao et~al.(2020)Cao, Fang, Liu, and Gong}]{fltrust}
Cao, X.; Fang, M.; Liu, J.; and Gong, N.~Z. 2020.
\newblock Fltrust: Byzantine-robust federated learning via trust bootstrapping.
\newblock \emph{arXiv preprint arXiv:2012.13995}.

\bibitem[{Fang and Chen(2023)}]{F3BA}
Fang, P.; and Chen, J. 2023.
\newblock On the vulnerability of backdoor defenses for federated learning.
\newblock In \emph{Proceedings of the AAAI Conference on Artificial Intelligence}, volume~37, 11800--11808.

\bibitem[{Fung, Yoon, and Beschastnikh(2018)}]{foolsgold}
Fung, C.; Yoon, C.~J.; and Beschastnikh, I. 2018.
\newblock Mitigating sybils in federated learning poisoning.
\newblock \emph{arXiv preprint arXiv:1808.04866}.

\bibitem[{Gu, Dolan-Gavitt, and Garg(2017)}]{badnets}
Gu, T.; Dolan-Gavitt, B.; and Garg, S. 2017.
\newblock Badnets: Identifying vulnerabilities in the machine learning model supply chain.
\newblock \emph{arXiv preprint arXiv:1708.06733}.

\bibitem[{Hard et~al.(2018)Hard, Rao, Mathews, Ramaswamy, Beaufays, Augenstein, Eichner, Kiddon, and Ramage}]{mobile}
Hard, A.; Rao, K.; Mathews, R.; Ramaswamy, S.; Beaufays, F.; Augenstein, S.; Eichner, H.; Kiddon, C.; and Ramage, D. 2018.
\newblock Federated learning for mobile keyboard prediction.
\newblock \emph{arXiv preprint arXiv:1811.03604}.

\bibitem[{He et~al.(2016)He, Zhang, Ren, and Sun}]{resnet18}
He, K.; Zhang, X.; Ren, S.; and Sun, J. 2016.
\newblock Deep residual learning for image recognition.
\newblock In \emph{Proceedings of the IEEE conference on computer vision and pattern recognition}, 770--778.

\bibitem[{Hsu, Qi, and Brown(2019)}]{Dirichlet}
Hsu, T.-M.~H.; Qi, H.; and Brown, M. 2019.
\newblock Measuring the effects of non-identical data distribution for federated visual classification.
\newblock \emph{arXiv preprint arXiv:1909.06335}.

\bibitem[{Kirkpatrick et~al.(2017)Kirkpatrick, Pascanu, Rabinowitz, Veness, Desjardins, Rusu, Milan, Quan, Ramalho, Grabska-Barwinska et~al.}]{Continual1}
Kirkpatrick, J.; Pascanu, R.; Rabinowitz, N.; Veness, J.; Desjardins, G.; Rusu, A.~A.; Milan, K.; Quan, J.; Ramalho, T.; Grabska-Barwinska, A.; et~al. 2017.
\newblock Overcoming catastrophic forgetting in neural networks.
\newblock \emph{Proceedings of the national academy of sciences}, 114(13): 3521--3526.

\bibitem[{Kone{\v{c}}n{\`y} et~al.(2016)Kone{\v{c}}n{\`y}, McMahan, Yu, Richt{\'a}rik, Suresh, and Bacon}]{2016federated}
Kone{\v{c}}n{\`y}, J.; McMahan, H.~B.; Yu, F.~X.; Richt{\'a}rik, P.; Suresh, A.~T.; and Bacon, D. 2016.
\newblock Federated learning: Strategies for improving communication efficiency.
\newblock \emph{arXiv preprint arXiv:1610.05492}.

\bibitem[{Krizhevsky, Hinton et~al.(2009)}]{cifar10}
Krizhevsky, A.; Hinton, G.; et~al. 2009.
\newblock Learning multiple layers of features from tiny images.

\bibitem[{Le and Yang(2015)}]{tiny-imagenet}
Le, Y.; and Yang, X. 2015.
\newblock Tiny imagenet visual recognition challenge.
\newblock \emph{CS 231N}, 7(7): 3.

\bibitem[{Li et~al.(2022{\natexlab{a}})Li, Bhagoji, Chen, Zheng, and Zhao}]{continual3}
Li, H.; Bhagoji, A.~N.; Chen, Y.; Zheng, H.; and Zhao, B.~Y. 2022{\natexlab{a}}.
\newblock On the Permanence of Backdoors in Evolving Models.
\newblock \emph{arXiv preprint arXiv:2206.04677}.

\bibitem[{Li et~al.(2021)Li, Meng, Ma, Du, Zhu, Pei, and Shen}]{smart-healthcare-system}
Li, J.; Meng, Y.; Ma, L.; Du, S.; Zhu, H.; Pei, Q.; and Shen, X. 2021.
\newblock A federated learning based privacy-preserving smart healthcare system.
\newblock \emph{IEEE Transactions on Industrial Informatics}, 18(3).

\bibitem[{Li et~al.(2022{\natexlab{b}})Li, Jiang, Li, and Xia}]{dataset—jieshao}
Li, Y.; Jiang, Y.; Li, Z.; and Xia, S.-T. 2022{\natexlab{b}}.
\newblock Backdoor learning: A survey.
\newblock \emph{IEEE transactions on neural networks and learning systems}, 35(1): 5--22.

\bibitem[{Lin et~al.(2020)Lin, Kong, Stich, and Jaggi}]{feddf}
Lin, T.; Kong, L.; Stich, S.~U.; and Jaggi, M. 2020.
\newblock Ensemble distillation for robust model fusion in federated learning.
\newblock \emph{Advances in neural information processing systems}, 33: 2351--2363.

\bibitem[{Liu et~al.(2024)Liu, Zhang, Feng, Yang, Xu, Man, and Yang}]{FCBA}
Liu, T.; Zhang, Y.; Feng, Z.; Yang, Z.; Xu, C.; Man, D.; and Yang, W. 2024.
\newblock Beyond traditional threats: A persistent backdoor attack on federated learning.
\newblock In \emph{Proceedings of the AAAI Conference on Artificial Intelligence}, volume~38, 21359--21367.

\bibitem[{Lyu et~al.(2022)Lyu, Yu, Ma, Chen, Sun, Zhao, Yang, and Yu}]{FL}
Lyu, L.; Yu, H.; Ma, X.; Chen, C.; Sun, L.; Zhao, J.; Yang, Q.; and Yu, P.~S. 2022.
\newblock Privacy and robustness in federated learning: Attacks and defenses.
\newblock \emph{IEEE transactions on neural networks and learning systems}, 35(7): 8726--8746.

\bibitem[{Mandt, Hoffman, and Blei(2017)}]{SGD}
Mandt, S.; Hoffman, M.~D.; and Blei, D.~M. 2017.
\newblock Stochastic gradient descent as approximate bayesian inference.
\newblock \emph{Journal of Machine Learning Research}, 18(134): 1--35.

\bibitem[{McMahan et~al.(2017)McMahan, Moore, Ramage, Hampson, and y~Arcas}]{Federated—Learning}
McMahan, B.; Moore, E.; Ramage, D.; Hampson, S.; and y~Arcas, B.~A. 2017.
\newblock Communication-efficient learning of deep networks from decentralized data.
\newblock In \emph{Artificial intelligence and statistics}, 1273--1282. PMLR.

\bibitem[{Nguyen et~al.(2022)Nguyen, Rieger, Chen, Yalame, M{\"o}llering, Fereidooni, Marchal, Miettinen, Mirhoseini, Zeitouni et~al.}]{flame}
Nguyen, T.~D.; Rieger, P.; Chen, H.; Yalame, H.; M{\"o}llering, H.; Fereidooni, H.; Marchal, S.; Miettinen, M.; Mirhoseini, A.; Zeitouni, S.; et~al. 2022.
\newblock $\{$FLAME$\}$: Taming backdoors in federated learning.
\newblock In \emph{31st USENIX security symposium (USENIX Security 22)}, 1415--1432.

\bibitem[{Parisi et~al.(2019)Parisi, Kemker, Part, Kanan, and Wermter}]{continual2}
Parisi, G.~I.; Kemker, R.; Part, J.~L.; Kanan, C.; and Wermter, S. 2019.
\newblock Continual lifelong learning with neural networks: A review.
\newblock \emph{Neural networks}, 113: 54--71.

\bibitem[{Pillutla, Kakade, and Harchaoui(2022)}]{rfa}
Pillutla, K.; Kakade, S.~M.; and Harchaoui, Z. 2022.
\newblock Robust aggregation for federated learning.
\newblock \emph{IEEE Transactions on Signal Processing}, 70: 1142--1154.

\bibitem[{Ren et~al.(2024)Ren, Zheng, Yang, Li, and Ma}]{shadow_attack}
Ren, Q.; Zheng, Y.; Yang, C.; Li, Y.; and Ma, J. 2024.
\newblock Shadow backdoor attack: Multi-intensity backdoor attack against federated learning.
\newblock \emph{Computers \& Security}, 139: 103740.

\bibitem[{Rodr{\'\i}guez-Barroso et~al.(2023)Rodr{\'\i}guez-Barroso, Jim{\'e}nez-L{\'o}pez, Luz{\'o}n, Herrera, and Mart{\'\i}nez-C{\'a}mara}]{Concepts-taxonomy}
Rodr{\'\i}guez-Barroso, N.; Jim{\'e}nez-L{\'o}pez, D.; Luz{\'o}n, M.~V.; Herrera, F.; and Mart{\'\i}nez-C{\'a}mara, E. 2023.
\newblock Survey on federated learning threats: Concepts, taxonomy on attacks and defences, experimental study and challenges.
\newblock \emph{Information Fusion}, 90: 148--173.

\bibitem[{Shukla et~al.(2025)Shukla, Rajkumar, Sinha, Esha, Elango, and Sampath}]{shukla2025federated}
Shukla, S.; Rajkumar, S.; Sinha, A.; Esha, M.; Elango, K.; and Sampath, V. 2025.
\newblock Federated learning with differential privacy for breast cancer diagnosis enabling secure data sharing and model integrity.
\newblock \emph{Scientific Reports}, 15(1): 13061.

\bibitem[{Simonyan and Zisserman(2014)}]{vgg16}
Simonyan, K.; and Zisserman, A. 2014.
\newblock Very deep convolutional networks for large-scale image recognition.
\newblock \emph{arXiv preprint arXiv:1409.1556}.

\bibitem[{Sun et~al.(2019)Sun, Kairouz, Suresh, and McMahan}]{dp}
Sun, Z.; Kairouz, P.; Suresh, A.~T.; and McMahan, H.~B. 2019.
\newblock Can you really backdoor federated learning?
\newblock \emph{arXiv preprint arXiv:1911.07963}.

\bibitem[{Wang, Quasim, and Yi(2025)}]{smart-healthcare}
Wang, J.; Quasim, M.~T.; and Yi, B. 2025.
\newblock Privacy-preserving heterogeneous multi-modal sensor data fusion via federated learning for smart healthcare.
\newblock \emph{Information Fusion}, 120: 103084.

\bibitem[{Xie et~al.(2019)Xie, Huang, Chen, and Li}]{DBA}
Xie, C.; Huang, K.; Chen, P.-Y.; and Li, B. 2019.
\newblock Dba: Distributed backdoor attacks against federated learning.
\newblock In \emph{International conference on learning representations}.

\bibitem[{Xie, Koyejo, and Gupta(2019)}]{zeno}
Xie, C.; Koyejo, S.; and Gupta, I. 2019.
\newblock Zeno: Distributed stochastic gradient descent with suspicion-based fault-tolerance.
\newblock In \emph{International conference on machine learning}, 6893--6901. PMLR.

\bibitem[{Yin et~al.(2018)Yin, Chen, Kannan, and Bartlett}]{trimmed_mean}
Yin, D.; Chen, Y.; Kannan, R.; and Bartlett, P. 2018.
\newblock Byzantine-robust distributed learning: Towards optimal statistical rates.
\newblock In \emph{International conference on machine learning}, 5650--5659. Pmlr.

\bibitem[{Yurochkin et~al.(2019)Yurochkin, Agarwal, Ghosh, Greenewald, Hoang, and Khazaeni}]{non-iid2}
Yurochkin, M.; Agarwal, M.; Ghosh, S.; Greenewald, K.; Hoang, N.; and Khazaeni, Y. 2019.
\newblock Bayesian nonparametric federated learning of neural networks.
\newblock In \emph{International conference on machine learning}, 7252--7261. PMLR.

\bibitem[{Zhang et~al.(2023)Zhang, Jia, Chen, Lin, and Wu}]{A3FL}
Zhang, H.; Jia, J.; Chen, J.; Lin, L.; and Wu, D. 2023.
\newblock A3fl: Adversarially adaptive backdoor attacks to federated learning.
\newblock \emph{Advances in neural information processing systems}, 36: 61213--61233.

\bibitem[{Zhang et~al.(2022)Zhang, Panda, Song, Yang, Mahoney, Mittal, Kannan, and Gonzalez}]{neurotoxin}
Zhang, Z.; Panda, A.; Song, L.; Yang, Y.; Mahoney, M.; Mittal, P.; Kannan, R.; and Gonzalez, J. 2022.
\newblock Neurotoxin: Durable backdoors in federated learning.
\newblock In \emph{International Conference on Machine Learning}, 26429--26446. PMLR.

\bibitem[{Zhu et~al.(2021)Zhu, Xu, Liu, and Jin}]{non-iid1}
Zhu, H.; Xu, J.; Liu, S.; and Jin, Y. 2021.
\newblock Federated learning on non-IID data: A survey.
\newblock \emph{Neurocomputing}, 465: 371--390.

\bibitem[{Zhuang et~al.(2023)Zhuang, Yu, Wang, Hua, Li, and Yuan}]{BC_layers}
Zhuang, H.; Yu, M.; Wang, H.; Hua, Y.; Li, J.; and Yuan, X. 2023.
\newblock Backdoor federated learning by poisoning backdoor-critical layers.
\newblock \emph{arXiv preprint arXiv:2308.04466}.

\end{thebibliography}
\end{document}